\documentclass[12pt]{article}
\usepackage{epsf}
\usepackage{alltt}
\usepackage{morefloats}
\usepackage{epsfig}
\def\largelinestretch{\renewcommand{\baselinestretch}{1.1}}
\largelinestretch\small\normalsize

\newcommand{\dvder}[2]{\frac {\displaystyle \partial #1}{\displaystyle \partial #2}}
\newcommand{\hder}[2]{\partial #1/\partial #2}
\newcommand{\dvrat}[2]{\frac {\displaystyle #1}{\displaystyle #2}}
\newcommand{\cfk}{\cos\phi_k}
\newcommand{\sfk}{\sin\phi_k}

\title{
\vspace*{-15mm}
{\normalsize \begin{tabular}[t]{ll}
\hspace{1cm}DESY 08--182 &    \\
\hspace{1cm}December 2008 &  \\
& \end{tabular}
  \hfill { } \\}
\vspace*{15mm}
{\bf Mathematical Framework\\
 for Fast and Rigorous Track Fit\\ 
for the ZEUS Detector}
}
 
\author{
       {\it Alexander Spiridonov}\footnote{
E-mail: Alexander.Spiridonov@desy.de 
\newline 
\hspace*{5mm} Permanent address: 
Institute of Theoretical and Experimental Physics, 117259 Moscow, Russia
}\\
        DESY }
\date{}
 
\begin{document}
\maketitle
\begin{abstract}
In this note we present 
a mathematical framework for a rigorous approach to a 
common track fit for trackers located in the inner region of the 
ZEUS detector. The approach makes use  
of the Kalman filter and offers a rigorous treatment of magnetic field 
inhomogeneity, multiple scattering and energy loss. 
We describe mathematical details
of the implementation of the Kalman filter technique
with a reduced amount of computations for 
a cylindrical drift chamber, barrel and forward silicon strip detectors 
and a forward straw drift chamber. Options with homogeneous and 
inhomogeneous field are discussed.
The fitting of tracks in one ZEUS event
takes about of 20\,ms on 
%a modern PC with a single core processor.
standard PC.
\end{abstract}

\clearpage
\tableofcontents
%page4
\clearpage
%=======================
 \section{Introduction}
%=======================
The ZEUS experiment~\cite{zeus} was 
operated 
%conducted
%a large multi-component detector 
at the electron-proton collider HERA at DESY until 2007.
%In collision mode the protons and electrons (positrons)
%had an energy 920 GeV and 27.5 GeV, respectively.
The ZEUS detector was a sophisticated, multi-component tool 
for studying particle reactions provided by 
electron-proton collisions with an energy 27.5 GeV and 920 GeV,respectively.
The inner tracking components of the ZEUS detector
were: the silicon strip Micro Vertex Detector~\cite{mvd} with 
barrel (BMVD) and forward (FMVD) parts; 
the Central Tracking Detector (CTD)~\cite{ctd}
consisting of the cylindrical 
%jet type 
drift chamber;
the Forward Tracking Device (FTD)~\cite{zeus}
and the forward Straw-Tube Tracker (STT)~\cite{stt}.
The MVD was located in the vicinity of  
interaction point, inside of the CTD.
%The FTD and STT followed MVD and CTD and covered 
%its acceptance in the forward direction.

The magnetic field in the central region of the ZEUS detector
was produced by a thin superconducting solenoid.
The field had a strength of 14.3 kGauss at the center and was
directed parallel to the proton beam.
The barrel MVD and CTD were located in the field 
which was almost homogeneous with a small radial component
far from the center.  
Forward trackers were placed outside of the solenoid or
close to its edge where the field is inhomogeneous.  

We consider a mathematical framework for a rigorous approach to a common 
track fit, which can be performed
with tracks including all inner tracking components
or with any combination of them.
The approach offers a rigorous treatment of field inhomogeneity,
multiple scattering and energy loss.
The track fitting procedure makes use 
of the Kalman filter technique and we discuss how to optimize computations
and make the fitting procedure fast. 

%======================= 
 \section{Overview of the tracker layout} 
%=======================
The ZEUS coordinate system is a right-handed Cartesian system, with the 
$z$--axis pointing in the proton beam direction (forward) and the $x$--axis 
pointing 
to the center of the HERA ring. The coordinate origin is at the nominal
interaction point. 

The barrel (BMVD) and forward (FMVD) section of the MVD
includes 600 and 112 sensors, respectively~\cite{mvd}.
A sensor is a silicon single-sided strip detector
with a readout pitch of $120\,\mu{\rm m}$ which includes five innermost
strips for capacitive charge division. 
The ZEUS MVD has 307,200 and 53,730 readout channels in the barrel
and forward sections, respectively.
 
The barrel section, centered at the interaction point, is about 63 cm long.
The silicon sensors are arranged
in three concentric cylindrical layers with radii about 5\,cm, 8\,cm 
and 12\,cm. 
Two back to back sensors in a layer provide measurements of
nominal $r-\phi$ and $z$ position.
The FMVD is composed of 
%four vertical planes extending 
four transverse disks of 14 wedges each, which extend
the angular
coverage down to $7^\circ$ from the beam line. 
Each wedge has two sensor layers separated by approximately  
8\,mm in $z$--direction.
%Sensors in two layers 
They
are mounted back to back,
such that the angle between strips is $2\times13^\circ$. 

The CTD~\cite{mvd} is a cylindrical drift chamber, with a sensitive 
volume approximately 2m in length and 0.4 (1.6m) in inner (outer) diameter. 
The CTD wires are arranged into nine 
%page5
concentric superlayers
numbered consecutively from the inside out.
The odd-numbered superlayers have sense wires running parallel
to the chamber axis (i.e. $z$--axis) while those in the even-numbered
superlayers have a $5^\circ$ stereo angle.
%Three-dinentional information is calculated using these small angles 
%stereo layers.
We denote sense wires
in corresponding superlayers as axial and stereo, respectively.
Each superlayer contains eight sense wire layers -- there are 4608
sense wires in total. 
%These eight sense wire are 
A set of eight sense wires is
surrounded by field wires,
%which divide 
azimuthally dividing a 
superlayer into cells of polygonal shape.
Each sense wire is read out by a flash ADC and, finally a drift distance
is evaluated for a hit. 
All axial wires in superlayer one and
the odd numbered wires in superlayer three and five (in total 704 wires)
are additionally equipped with the z-by-timing system, which
measures $z$ position of a hit.

The STT uses straw drift chambers 
%with capton tube of 7.5mm diameter and of varying length 
with 7.5\,mm diameter capton tubes of varying length
from 20\,cm to 75\,cm. 
There are in total 10,944 wires 
%in 48 sectors of wedge shape.
in 48 wedge shaped sectors. 
%Wedge covers an azimuthal angle region 
Each wedge covers an azimuthal span
of $60^\circ$. Each sector consists 
of 3 layers of straws perpendicular to the $z$--axis. 
%Sectors are rotated against each other to cover the total range of azimuthal angles. 
%Eight z stations each have six sectors covering the full azimuthal range.   
A track crossing the STT nominally delivers 24 drift time measurements. 

%======================= 
 \section{Track Models and Likelihood Functions 
in a Multi-Component Tracker} 
%=======================
The likelihood function of a track measurement
has a meaning regardless of the details of any fitting method.
The maximum-likelihood estimator is
{\it efficient} in the sense 
that no other unbiased estimator has smaller variances.
%that there is no estimator with smaller variances. 
%Usage of a track model similar to those
%which is suitable for the likelihood function, may give a chance to perform an
%efficient parameter estimation with a given method of track fit.
A track model which is appropriate for the likelihood function,
together with a given method of track fit,
may produce an efficient estimate of parameters.
A general point of view of the information delivered by a tracker
can help to interpret behavior of variances of fitted parameters
and hit residuals.     
 
We can model a multi-component tracker by a set of track detecting elements
and intermediate blocks of passive material, which are located 
in a static magnetic field.
Track parameters in the detector element $k$ are described by a vector
$x_k$. For the case of a three-dimensional fit, 
the dimension of the vector,$x_k$, is 5.
The track measurement in the tracker element $k$, 
i.e. the $k^{th}$ hit, is a vector denoted by $m_k$.
In general $m_k$ is the vector with its dimension corresponding to that of
the tracking element. For example, $m_k$ has only 1 coordinate for
a silicon strip of the MVD, a drift tube of the STT
or a stereo wire of the CTD 
%and 2 coordinates for an axial wire of the CTD with 
%an additional $z$ measurement of the track coordinate.
and 2 coordinates (drift time and $z$ position of a hit)
for an axial wire of the CTD which is
additionally equipped with the z-by-timing system.
The measurement error can be described by the covariance matrix $V_k$.
We approximate
the probability (density) of the measurement $m_k$ given the vector 
of track parameters $x_k$ 
\begin{equation}
P( m_k | x_k ) = G( m_k | \langle m_k \rangle; V_k)
\label{pmeas}
\end{equation}
%page6
by a Gaussian function
%, $G( m_k | \langle m_k \rangle; V_k)$,
with the mean value $\langle m_k \rangle$ and covariance matrix $V_k$:
\begin{equation}
 G( m_k | \langle m_k \rangle; V_k) = 
C(V_k)\,{\rm exp}\, 
\Big\{ -\frac{1}{2} {(m_k-\langle m_k \rangle)}^T
V_k^{-1} (m_k-\langle m_k \rangle) \Big\},
\label{mgauss}
\end{equation}
where $C(V_k)$ is a normalization constant. 
An operator $H_k$ projects the actual vector $x_k$ into the space of
measurement:
\begin{equation}
 \langle m_k \rangle = H_k\,x_k.
\label{ex}
\end{equation}

Suppose that we are interested in the track parameters at the
beginning of track, $x_1$. The likelihood function takes the form of
a product:
\begin{equation}
L(m_1, m_2, ..., m_N | x_1) = P(x_2, ..., x_N | x_1) \, \cdot \,
\prod_{k=1}^N G( m_k | H_k\,x_k; V_k).
\label{lx1}
\end{equation}
The first term  
is the probability for a particle to pass through the points  
$x_2, ..., x_N$ given the parameters $x_1$ at the beginning
and the second one is  the probability to obtain the 
measurements, $m_1, ..., m_N$, while measuring the points in the space  
of track parameters $x_1, ..., x_N$ of the real (not the mean) trajectory.
The probability, $P(x_2, ..., x_N | x_1)$, can be approximated 
by a Gaussian distribution
\begin{equation}
P(x_2, ..., x_N | x_1) = 
G( x_2, ...,x_N | \langle x_2(x_1)\rangle, ..., \langle x_N(x_1)\rangle ; 
\Sigma(x_1)).
\label{pxnx2}
\end{equation}
The mean trajectory is defined as:
\begin{equation}
\langle x_k(x_1)\rangle = {\cal F}_k x_1,
\label{x1swim}
\end{equation}
where the operator ${\cal F}_k$ swims track parameters $x_1$ into 
the detector element $k$.
The track model may be described as
a continuous curve for the mean trajectory
%which is  
%a solution of equations of a particle motion in a static magnetic field
%with initial parameters $x_1$. 
with fluctuations of actual parameters $x_k$ with respect to the 
mean trajectory,
\begin{equation} 
{\cal D}_k(x_1) = x_k-{\cal F}_k x_1. 
\label{dx1} 
\end{equation} 
The fluctuation, ${\cal D}_k(x_1)$, accumulates the effect of 
multiple scattering on the pass 
from the beginning
of the track to the given element. Vectors $\{{\cal D}_k(x_1)\}$ are correlated  
%correlated for differnt $k$  
and, therefore, matrix $\Sigma(x_1)$ has dense structure (many non-zero
elements). We can combine Gaussian functions from (\ref{lx1}) and
(\ref{pxnx2}): 
\begin{equation}
L(m_1, m_2, ..., m_N | x_1) = 
G(m_1, m_2, ..., m_N | H_1x_1, H_2{\cal F}_2x_1, ..., H_N{\cal F}_Nx_1; 
{\cal M}(x_1)),
\label{lx12g}
\end{equation}
where the non-diagonal covariance matrix ${\cal M}(x_1)$ 
has dimension equal to the sum of dimensions of all measurements $\{m_k\}$.
The dimension of the ${\cal M}$ may be of order $10^2$ for 
modern tracking detectors.
Maximization of the likelihood function
of Gaussian type, i.e. least square fitting
with large non-diagonal covariance matrix ${\cal M}$, 
requires a lot of computations, although not more than 5 parameters are
fitted. Because of large computing time,
the model is not convenient for a track fitting in a multi-component 
tracker. But the model includes a small number of fitted parameters,
and is suitable for 
%page7
a subsequent update of detector alignment 
parameters~\cite{mvdalign}, where an expansion of
hit residuals w.r.t. fitted parameters is needed. 

A charged particle traversing a medium can be described by 
%The process of multiple scattering is 
a stochastic process with the Markov 
property and, therefore, the conditional probability distribution of future
states 
%of the process 
depends only upon the present state and not on any past
states. The probability function
for a particle to pass through the points  
$x2, ..., x_N$ 
%if it began motion with the parameters $x_1$ 
in (\ref{pxnx2}) can be rewritten as:
\begin{equation}
P(x_2, ..., x_N | x_1) = \prod_{k=2}^N P(x_k | x_{k-1} ) = 
\prod_{k=2}^N G( x_k | \langle x_k(x_{k-1}) \rangle; Q_k(x_{k-1})).
\label{pmarkov}
\end{equation}
%In the latter expession,
We approximate the conditional probability (density),
$P(x_k | x_{k-1} )$, for track parameters $x_k$, given
the parameters in the previous state $x_{k-1}$, by
the Gaussian distribution, 
$G( x_k | \langle x_k(x_{k-1}) \rangle; Q_k(x_{k-1}))$
with the mean $\langle x_k(x_{k-1}) \rangle$ and covariance
matrix $Q_k(x_{k-1})$. 
The mean trajectory in the tracking element $k$ is   
\begin{equation}
\langle x_k(x_{k-1})\rangle = F_k x_{k-1}.
\label{xkswim}
\end{equation}
%given the vector of track parameters in the previous element $x_{k-1}$. 
The operator $F_k$ 
%in (\ref{xkswim}) 
swims track parameters $x_{k-1}$ into 
the detector element $k$ according to the equations of motion.

Suppose that we are interested in track parameters 
in all points of track measurement, i.e. $x_1, x_2, ..., x_N$.
The likelihood function takes a form: 
\begin{equation}
L(m_1, ..., m_N | x_1, ..., x_N ) = G( m_1 | H_1\,x_1; V_1) \cdot
\prod_{k=2}^N G( x_k | F_k\,x_{k-1}; Q_k)\,G( m_k | H_k\,x_k; V_k),
\label{lxn}
\end{equation}
with Gaussian functions
\begin{equation}
 G( m_k | H_k\,x_k; V_k) = 
C(V_k)\,{\rm exp}\, 
\Big\{ -\frac{1}{2} {(m_k - H_k\,x_k)}^T
V_k^{-1} (m_k - H_k\,x_k) \Big\}
\label{mkgauss}
\end{equation}
and 
\begin{equation}
 G( x_k | F_k\,x_{k-1}; Q_k) = 
C(Q_k)\,{\rm exp}\, 
\Big\{ -\frac{1}{2} {(x_k - F_k\,x_{k-1})}^T
Q_k^{-1} (x_k - F_k\,x_{k-1}) \Big\},
\label{xkgauss}
\end{equation}
where $C(V_k)$ and $C(Q_k)$ are normalization constants.

The model for the total track is not a continuous curve, but consists of
$N-1$ continuous segments. 
A variation of track parameters in the point of discontinuity 
\begin{equation} 
\delta_k = x_k - F_k x_{k-1} 
\label{dxk} 
\end{equation} 
describes the effect of multiple scattering on the pass from the 
the previous element $k-1$ to the element $k$.
%and where we may expect an additional material. 
Vectors $\{\delta_k\}$  are uncorrelated.
A spread of the $\delta_k$ is defined by the covariance matrix $Q_k$.

The maximum likelihood estimation of parameters $\{{x_k}\}$ 
satisfies the system of
%is given by the solution of 
equations 
\begin{equation}
\left\{ \frac {\partial\,(-{\rm ln}\,L)}{\partial x_k^T}\,=\,0 \right\}.
\label{dldx} 
\end{equation} 
%page8
If operators $F_k$ and $H_k$ are non-linear (e.g. in magnetic field)
then the latter equations are non-linear also.
The problem can be solved
iteratively using the well known method of linearization of
operations (\ref{ex}) and (\ref{xkswim}). 
Anyhow we can regard the functional, $\partial(-{\rm ln}\,L) / \partial x_k^T$,
as a linear form w.r.t. vectors of estimated parameters $\{x_k\}$.   
The vector, $x_k$, associates
in (\ref{lxn}) only
with vectors in neighboring data points $k-1$ and $k+1$  and,
therefore, the linear form $\partial(-{\rm ln}\,L) / \partial x_k^T$
includes only 3 terms with vectors $x_{k-1}, x_k$ and 
$x_{k+1}$\footnote{Differentiating the latter linear form with respect
to $x_i$ isolates a coefficient in a corresponding linear term.}, respectively.
Finally, the system (\ref{dldx}) looks as 
\begin{equation}
\label{ixr}
\left( \begin{array}{ccccccc}
%   1        2      3        4        5        6     7
\,I_{11} & I_{12} &      &      &          &      &    \\
\,I_{21} & I_{22} &I_{23}&      &          &      &    \\
         & I_{32} &I_{33}&I_{34}&          &      &    \\
         &        & ...  & ...  & ...      &      &    \\  
         &        &      & ...  & ...      & ...  &    \\  
         &        &      &      &I_{N-1\,N}&I_{NN}&    \\
\end{array} \right)
\left( \begin{array}{c}
x_1 \\
x_2 \\
x_3 \\
... \\
... \\
x_N \\ 
\end{array}
\right)
\,=\,
\left( \begin{array}{c}
r_1 \\
r_2 \\
r_3 \\
... \\
... \\
r_N \\ 
\end{array}
\right),
\end{equation}
where submatrices related to points $i, j$,
\[
I_{ij} = \frac {\partial^2 (-{\rm ln}\,L)} {\partial x_i^T \partial x_j},
\]
are parts of the information matrix.
%,${\bf I}$.
The sparse (with many zero elements), 
band structure of the information matrix can
be exploited to reduce computations drastically.
%This can be achieved 
%either by the usage of dedicated algorithm of matrix 
%inversion~\cite{multpl} or like in the broken lines fit~\cite{blobel},
%by the matrix (Cholesky) decomposition 
This can be achieved by using either a dedicated algorithm of matrix
inversion~\cite{multpl}, or else 
%-- like in the broken lines fit~\cite{blobel} -- 
(e.g. in the broken lines fit~\cite{blobel}) 
by the matrix (Cholesky) decomposition
into a unit triangle, $U$, 
and a diagonal, $D$, 
%matrices
matrix
\[
U\,D\,U^T {\bf x} = {\bf r}
\]
%and searching of a solution in two steps:
which requires two steps to solve for ${\bf x}$:
\[
U\,{\bf y} = {\bf r}\,\,\,\,\,{\rm and}\,\,\,\,\,D\,U^T {\bf x} = {\bf y}.
\]
The track model based on relations 
(\ref{xkswim} -- \ref{dxk}) is well suited also for an 
implementation of the progressive track fit by the method~\cite{bill}
or for the application of the Kalman filter formalism~\cite{fruhw}.
Both methods are rather economical regarding computing time because
they include operations with matrices of maximal size 5 by 5 for
each hit.
     
%======================= 
 \section{Application of the Kalman filter technique  to track fitting} 
%=======================
In~\cite{fruhw} it was shown that an appropriate mathematical framework
for the iterative procedure of track fitting is the theory of
linear filtering, in particular the Kalman filter~\cite{kalman}.
%Filtering deals with the analysis of linear dynamic systems~\cite{lfilter}.
%We can regard the movement of a particle
%from one tracker element to the next 
%as an evolution of a dynamic system  from one to the next state
%and apply filtering techniques to track fitting. 
%We can identify the state vector of the 
%dynamic system with the vector $x$ of track parameters
%uniquely describing the track in each point of its trajectory. 
To consider the mathematical framework of a Kalman filter, we try to follow
the notation used in~\cite{mankel}.
%~\cite{fruhw} and in the review
%a clearly structured note 
In the following we describe a case with a linear system and
a non-linear system 
will be discussed at Subsec.~\ref{nonlinear}. 

%page9
%======================= 
 \subsection{Linear Model} 
%=======================
The Kalman filter proceeds progressively  from one measurement
to the next and improves the knowledge about the particle trajectory by
updating the track parameters with each new measurement.
The system state vector (track parameters) after inclusion of $k-1$ measurements
is denoted by $\tilde{x}_{k-1}$, and its covariance matrix by $C_{k-1}$.
The state vector and its covariance matrix are propagated to the
location of the next measurement with the {\it prediction equations}:
\begin{equation}
\tilde{x}^{k-1}_k  = F_k \tilde{x}_{k-1},
\label{fxk1}  
\end{equation}  
and 
\begin{equation} 
\label{ck1}  
C^{k-1}_k = F_k C_{k-1} F^T_k + Q_k,
\end{equation}  
where $F_k$ is the transport matrix and 
$Q_k$ denotes the covariance matrix of the {\it process noise},
which occurs due to 
the random perturbation of the particle's trajectory.

The measurement of the vector $\tilde{x}^{k-1}_k$ and its covariance matrix
are denoted by $m_k$ and $V_k$, respectively. 
The {\it expected} measurement $m_k$ is described by the projection matrix
$H_k$.   
The estimated residuals are
\begin{equation}  
r^{k-1}_k = m_k - H_k\,\tilde{x}^{k-1}_k
\label{frk}  
\end{equation}  
and its covariance matrix become:
\begin{equation}
R^{k-1}_k = V_k + H_k\,C^{k-1}_k\,H_k^T.
\label{Crk}   
\end{equation}   
The updating of the system state vector after inclusion of
the measurement $k$ is defined by the {\it filter equations}:
\begin{equation}
\begin{array}{ll}
K_k = C^{k-1}_k\,H_k^T\,(R^{k-1}_k)^{-1}, &  \\
\tilde{x}_k = \tilde{x}^{k-1}_k + K_k\,r^{k-1}_k, &\\
C_k = (1 - K_k\,H_k)\,C^{k-1}_k, &
\end{array}
\label{kxc} 
\end{equation}
with the filtered residuals and its covariance matrix 
\begin{equation}  
\begin{array}{ll}  
r_k = (1 - H_k\,K_k)\,r^{k-1}_k, &
~~~~~~~~R_k = (1 - H_k\,K_k)\,V_k
= V_k - H_k C_k H_k^T. \\
\end{array}  
\label{rrk}  
\end{equation}  
The matrix, $K_k$, is called the {\it filtering (gain ) matrix}.
The filtered state vector is pulled towards the measurement and,
therefore the quadratic mean of the filtered residual is smaller than 
the measurement error.
The $\chi^2$ increment after the filtering of the state vector is given by:
\[\chi_k^2 = r^T_k\,R^{-1}_k\,r_k.~~~~~~~~~~~~~\] 

The track parameters after the filtering procedure are known with
optimal precision only at the last point of the fit. 
The smoothing part of the Kalman filter is a very useful complement,
which solves the problem of optimal parameter estimation
at every point of the trajectory.
The smoothing is also a recursive procedure which proceeds step by step
%page10
in the direction opposite to that of the filter with the
{\it smoother equations}:
\begin{equation}
\begin{array}{l}
A_k = C_k\,F^T_{k+1}\,(C^k_{k+1})^{-1}, \\ 
\tilde{x}^n_k =  \tilde{x}_k + A_k (\tilde{x}^n_{k+1} - \tilde{x}^k_{k+1}), \\
C^n_k = C_k + A_k (C^n_{k+1} - C^k_{k+1}) A^T_k, \\
r^n_k = m_k - H_k \tilde{x}^n_k, \\
R^n_k = R_k - H_k A_k (C^n_{k+1} - C^k_{k+1}) A^T_k H^T_k 
      = V_k - H_k C^n_k H^T_k.
\end{array}  
\label{smooth}  
\end{equation}
The smoothed state vector, $\tilde{x}^n_k$, is more precise, 
because it includes information from all measurements.
%The variance of the filtered state vector, $C_k$,
%is bigger than the variance of the smoothed state vector, $C^n_k$.
The variance of the smoothed state vector, $C^n_k$,
is smaller than the variance of the filtered state vector, $C_k$.
%The smoothed state vector is pulled toward the true value and, 
%therefore the 
%The mean squared smoothed residual is closer
%to the measurement error (detector resolution) than the filtered one. 
The quadratic mean of the smoothed residual is closer
to the measurement error (detector resolution) than the filtered one. 

%======================= 
 \subsection{Non--linear Model} 
%=======================
\label{nonlinear}
%In reality the 
A particle's motion in a detector with magnetic field is
a nonlinear process.
In case of a non-linear system, we have to replace the
transport, $F_k$,  and projection, $H_k$,
matrices in (\ref{fxk1}) and (\ref{frk}), respectively, by exact non-linear
functions:    
\begin{equation} 
\begin{array}{ll} 
\label{exac}
\tilde{x}^{k-1}_k  = f_k (\tilde{x}_{k-1}), &
~~~~~~~~~~~~~r^{k-1}_k = m_k - h_k(\tilde{x}^{k-1}_k).  \\
\label{xk1rk} 
\end{array} 
\end{equation} 
Jacobian matrices of these functions (Jacobians in the following)
\begin{equation} 
\label{rk}
\frac {\partial (f_k)} {\partial (\tilde{x}_{k-1})},  
~~~~~~~~~~~~~\frac {\partial (h_k)} {\partial (\tilde{x}^{k-1}_k)}  \\
\label{jakob} 
\end{equation} 
will be used in equations for covariance matrix propagation
(\ref{ck1}) and (\ref{Crk}) instead of $F_k$ and $H_k$, respectively.
%For non-linear systems, the 
In practice, estimation
with Kalman filter for a non-linear system shows properties similar to those of
maximum-likelihood estimation:   
\begin{itemize}
\item 
%The estimator is  asymptotically unbiased, i.e. the estimator
%approaches the true parameter as the number of measurements increases.
The estimator is  asymptotically unbiased, i.e. its bias tends to zero
as the number of measurements increases.
\item
%The estimator is asymptotically Gaussian, i.e. 
The distribution of deviations of estimated parameters from
true values approaches a Gaussian distribution also asymptotically,
i.e for sufficiently large number of measurements. 
\end{itemize}

%======================= 
 \section{Particle Motion in a Static Magnetic Field} 
%=======================
The equation of motion of a particle with momentum $\vec{p}$ 
(velocity $\vec{v}$) 
and charge $Q$ in a static magnetic field $\vec{B}$ is:
\begin{equation}
\frac{d\,\vec{p}}{d\,t} = \kappa \cdot Q \cdot \vec{v} \times \vec{B},
\label{dpdt}
\end{equation}
%page11
where coordinates $x,\,y,\,z$ are in cm, $p$ is in GeV/c,  
the magnetic field $B$ is in kGauss, and parameter $\kappa$
is equal:
\[  \begin{array}{l}
\kappa = 0.000299792458\,\,(\rm{GeV/c})\,kG^{-1}\,cm^{-1}. \;\; \\
\end{array}   \]
The distance along the trajectory of a particle 
(path length) is given by: 
\[
s = |\vec{v}| \cdot t.
\]
The unitary vector $\vec{n}$ pointing along the direction of the trajectory
is:
\begin{equation} 
\vec{n} = \frac{d\,\vec{x}}{d\,s}.
\label{n} 
\end{equation} 
Equation~(\ref{dpdt}) can be rewritten as:
\begin{equation}
\frac{d\,\vec{n}}{d\,s} = \kappa \cdot \frac{Q}{|\vec{p}|}
\cdot \vec{n} \times \vec{B} = \kappa \cdot q  
\cdot \vec{n} \times \vec{B},
\label{dnds}
\end{equation}
where $q = Q/|\vec{p}|$.  
The latter equation combined with Eq.~(\ref{n}) gives a system of
linear differential equations:
\begin{equation}
\begin{array}{lll}  
d\,x\,/\,d\,s & = & n_x,\\
d\,y\,/\,d\,s & = & n_y,\\
d\,z\,/\,d\,s & = & n_z,\\
d\,n_x\,/\,d\,s & = & \omega_z \cdot n_y - \omega_y \cdot n_z,\\
d\,n_y\,/\,d\,s & = & \omega_x \cdot n_z - \omega_z \cdot n_x,\\
d\,n_z\,/\,d\,s & = & \omega_y \cdot n_x - \omega_x \cdot n_y,\\
q & = & {\rm const},
\end{array}
\label{dxds}    
\end{equation}
where $\omega_i(s) = \kappa \cdot q \cdot B_i(\vec{x}(s))$.
%The latter system of equation can be used for both cases when 
%the momentum $\vec{p}$ is either parallel or orthogonal
%to the $z$--axis. 
%Components of the vector $\vec{n}$ are constrained by 
%\begin{equation} 
%\label{nxnynz} 
%n_x^2 + n_y^2 + n_z^2 = 1. 
%\end{equation}
%and the functions $n_z(s)$ and $d\,n_z(s)\,/\,d\,s$ and one 
%equation can be excluded from system~(\ref{dxds}) using eq.~\ref{nxnynz}. 
%But is not clear that a solving of the reduced system
%is faster, because additional calculations of square roots
%are needed.
%At least, in GEANT package~\cite{geant} the system of kinematical 
%equations (\ref{dxds}) are implemented 
%in the routine \verb+GRKUTA+
%to transport a particle in the magnetic field by 
%using Runge-Kutta method for numerical solving of the equations.

%======================= 
 \section{Multiple Scattering and Energy Loss} 
%=======================
%In the ZEUS inner tracker detectors designed
%to keep minimal amount of material,
The ZEUS inner tracking detectors were designed using minimal material.
We take account of the effect of multiple scattering 
in the approximation of thin scatterers.
Multiple scattering after traversing
a material of small thickness, $l$, results in the perturbation   
of angles and coordinates, but the effect on the latter
has an additional order of smallness $o(l)$ and can be neglected. 
The deflection of the particle momentum $\vec{p}$ due to
multiple scattering is decomposed into deflections in two orthogonal planes.
%which are also orthogonal to $\vec{p}$.
We define two unit vectors $\vec{n_1},\,\vec{n_2}$ which
in combination with $\vec{n}$ form a right-handed Cartesian system:
\begin{equation}
\label{n1n2}
\vec{n_1} = 
\frac{\displaystyle \vec{e_z}\times\vec{n}}
{\displaystyle |\vec{e_z}\times\vec{n}|}
= \frac{\displaystyle 1}{\displaystyle n_t}\left( \begin{array}{c}
-n_y \\
n_x \\
0 \\
\end{array}
\right),\,\,
\vec{n_2} = \vec{n_1}\times\vec{n} = \frac{\displaystyle 1}{\displaystyle n_t}
\left( \begin{array}{c}
n_x \cdot n_z \\
n_y \cdot n_z \\
-n_t^2 \\
\end{array}
\right),
\,{\rm with}\,n_t = \sqrt{n_x^2 + n_y^2}.
\end{equation}
The direction of the momentum after the scattering is:
\begin{equation} 
\label{nprim} 
\vec{n}\,' = \vec{n} + \theta_1\cdot\vec{n_1} + \theta_2\cdot\vec{n_2},
\end{equation}  
%page12
where $\theta_1,\,\theta_2$ are random variables with
\begin{equation}
\label{thetams}
<\,\theta_{1,2}\,> = 0,\,\, 
{\rm var}\,(\theta_{1,2})\,=\,\theta_{ms}^2,\,\, 
{\rm cov}(\theta_1,\theta_2) = 0.
\end{equation}
Here $\theta_{ms}$ is the well-known Moli\'{e}re theory
expression for RMS of the deflection angle 
of a charged particle traversing a medium~\cite{pdg}
\begin{equation} 
\label{sms}
\theta_{ms}(t/X_0) = \frac{\displaystyle 13.6\,{\rm MeV}}{\beta c p}
\sqrt{t\,/X_0}\,\,\left[1 + 0.038\,{\ln (t/X_0)}\,\right],
\end{equation}  
where $t/X_0$ is the material thickness in radiation
lengths, which has to account for the track inclination:
\begin{equation}
t = l \cdot \sqrt{1 + {(n_x/n_z)}^2 + {(n_y/n_z)}^2 }.
\label{tvsl} 
\end{equation}
We rewrite Eq.~(\ref{nprim}) for the deflection of components: 
\begin{equation}
\label{varn}
\delta \vec{n}\, = \left( \begin{array}{c} 
\delta n_x \\ \delta n_y \\ \delta n_z \\ \end{array} \right)
= \theta_1 \left( \begin{array}{c}
-n_y\,/\,n_t \\
~~n_x\,/\,n_t \\
0 \\
\end{array}
\right)\,\,+\,\,\theta_2
\left( \begin{array}{c}
n_x \cdot n_z\,/\,n_t \\
n_y \cdot n_z\,/\,n_t \\
-n_t \\
\end{array}
\right).
\end{equation}
Taking into account Eqs.~(\ref{thetams}), we derive:
\begin{equation}
\label{covms}
\begin{array}{ll}
<\,\vec{n}\,'\,> & =\,\,\, \vec{n}, \\
{\rm var}\,(n_x') & =\,\,\, \theta_{ms}^2\,( n_y^2 + n_x^2 n_z^2 )/n_t^2,\\
{\rm var}\,(n_y') & =\,\,\,\theta_{ms}^2\,( n_x^2 + n_y^2 n_z^2 )/n_t^2,\\
{\rm var}\,(n_z') & =\,\,\,\theta_{ms}^2\,n_t^2,\\
{\rm cov}\,(n_x',n_y') & = \,\,\,\theta_{ms}^2\,n_x\,n_y
(n_z^2 - 1)/n_t^2,\\
{\rm cov}\,(n_x',n_z') & = \,\,\,-\theta_{ms}^2\,n_x\,n_z,\\
{\rm cov}\,(n_y',n_z') & = \,\,\,-\theta_{ms}^2\,n_y\,n_z.
\end{array}
\end{equation}
%A matrix decribing variation of the state vector XXXXXXXX
%due to multiple scattering is equal:
%\begin{equation}
%\label{Qms}
%Q_{ms}\,\,=\,\,\left( 
%\begin{array}{ccccc}
%~~0~~&~~0~~&~~0~~&~~0~~&~~0~~\\
%0 & 0 & 0 & 0 & 0\\ 
%0 & 0 & {\rm var}\,(n_x') & {\rm cov}\,(n_x',n_y') & 0\\ 
%0 & 0 & {\rm cov}\,(n_x',n_y') & {\rm var}\,(n_y') & 0\\ 
%0 & 0 & 0 & 0 & 0\\
%\end{array}
%\right).
%\end{equation}
An ionization energy loss is regarded as a deterministic correction
to a track energy.
% when traversing a material. 
In the approximation of thin scatterer,
track energy, $E$, after the traversal of a material is: 
\begin{equation}
E' = E - (dE/dx)_{ion} \cdot t, 
\label{dedx}   
\end{equation}   
where $(dE/dx)_{ion}$ is the mean rate of ionization energy loss
in the material.
%=================================== 
 \section{Specifics of Kalman Filter Implementation
for the ZEUS Inner Trackers} 
%===================================
Seven equations (\ref{dxds}) describe a particle motion in a magnetic
field, 
although five parameters suffice to define 
the trajectory at any point. 
A suitable track parameterization may depend 
on the detector geometry and field shape.
The magnetic field in the central part of the ZEUS
detector is directed parallel to the $z$--axis.
For the large part of the MVD the field is almost homogeneous
with only a small radial component 
%page13
($<1\%$ at the edge of the BMVD).
For the most forward parts of the CTD and FMVD the inhomogeneity is larger,
with reduction of the axial component by 8\% and increasing of 
the radial component up to 15\%.
The STT detector is located outside the superconducting
solenoid where the field is inhomogeneous.
We choose a different way to proceed depending on 
the polar angle,$\theta$, of a track ($\tan\theta = p_t/p_z$):
\begin{itemize}
\item
we use an option with inhomogeneous field for ``forward'' tracks 
$(0 < \theta < 60^\circ)$;
\item
a homogeneous field model is used for ``central'' tracks 
$(60^\circ < \theta < 120^\circ)$;
\item
an inhomogeneous field is used also for ``rear'' tracks 
$(120^\circ < \theta < 180^\circ)$.
\end{itemize}
The set of measurements, $\{m_k\}$, with its covariance matrices,
$\{V_k\}$, and the map of magnetic field, $\vec{B}$, are 
%known as 
input for the track fit.
To develop 
a mathematical framework for Kalman filter implementation we have to make
the following steps:
\begin{itemize}
\item
Select a convenient parameterization of the state vector, $x_k$.
\item
Find a solution of the prediction equations, $f_k(x_{k-1})$,
and a function to project the vector $x_k$ to the measurement, 
$h_k(x_k)$. 
\item 
Obtain Jacobians of latter functions
\[
\frac {\partial (f_k)} {\partial ({x}_{k-1})},  
~~~\frac {\partial (h_k)} {\partial ({x}_k)}.~~~~~~~~~~~~~~~~~~~~~~~~~~~~~~~~~~~~~~~~\\
\]
\item
Define covariance matrix of the process noise, $Q_k$.
\end{itemize}
%======================= 
 \section{Cylindrical Parameterization for central tracks 
%in a Homogeneous Magnetic Field
}
%=======================
The magnetic field at the central region of the ZEUS superconducting
solenoid is nearly parallel to the $z$--axis ($B_x,B_y \approx 0$)
and has almost constant strength.  
Therefore we approximate it as homogeneous
on the path from one point to the next.
The system of equation (\ref{dxds}) looks as
\begin{equation}
\begin{array}{lll}  
d\,x\,/\,d\,s & = & n_x,~~~~~~~~~~~~~~~~~~~~~~~~~~~~~~~~~~~~~~~~~~~~~\\
d\,y\,/\,d\,s & = & n_y,\\
d\,z\,/\,d\,s & = & n_z,\\
d\,n_x\,/\,d\,s & = & \omega_z \cdot n_y \\
d\,n_y\,/\,d\,s & = & - \omega_z \cdot n_x,\\
d\,n_z\,/\,d\,s & = & 0\\
q & = & {\rm const},
\end{array}
\label{soldxds}    
\end{equation}
%page14
where $\omega_z = \kappa \cdot q \cdot B_z$.
The component $n_z$ is constant and the angle (azimuthal), $\phi$, 
of the track direction with the $x$--axis depends linearly on $s$:
\begin{equation}
\begin{array}{lll}
\phi(s) & = & \phi_0 - \omega_z  s,~~~~~~~~~~~~~~~~~~~~~~~~~~~~~~~~~~~~~~~~~~\\
n_x(s) & = & n_t\,\cos (\phi_0 - \omega_z s),\\
n_y(s) & = & n_t\,\sin (\phi_0 - \omega_z s),\\
n_z(s) & = & n_{z0}, \\
\end{array}
\label{helix}    
\end{equation}
where $\phi_0,  n_{z0}$ are initial values at $s=0$.
A pair of conserved quantities can be derived from (\ref{soldxds}):
%\begin{equation}
%\begin{array}{lll}
% d/ds\,( x(s) + \frac{1}{\omega}\,n_y(s) ) & = & 0,\\
% d/ds\,( y(s) - \frac{1}{\omega}\,n_x(s) ) & = & 0,\\
%\end{array} 
%\label{conserv}     
%\end{equation} 
%which are constant along the trajectory:
\begin{equation}
\begin{array}{lll}
\displaystyle
x(s) + \frac{1}{\omega_z}\,n_y(s) & = & 
\displaystyle
x_0 + \frac{1}{\omega_z}\,n_{y0},~~~~~~~~~~~~~~~~~~~~~~~~\\
\displaystyle
y(s) - \frac{1}{\omega_z}\,n_x(s) & = & 
\displaystyle
y_0 - \frac{1}{\omega_z}\,n_{x0},\\
\end{array} 
\label{conserv1}     
\end{equation} 
with initial values, $x_0, y_0, n_{x0}, n_{y0}$.
Coordinates can by expressed  via the track direction:
\begin{equation}
\begin{array}{l}\displaystyle
x(s) = x_0 - \frac{1}{\omega_z}\,n_y(s) + \frac{1}{\omega_z}\,n_{y0},~~~~~~~~~~~~~~~~~~~~~~~~~~~ \\
\displaystyle
y(s) = y_0 + \frac{1}{\omega_z}\,n_x(s) - \frac{1}{\omega_z}\,n_{x0}. \\
\end{array} 
\label{xnx}     
\end{equation} 
In a homogeneous field, the particle trajectory is a  helix.
For the  case of axial (cylindrical) symmetry,
a natural replacement of particle coordinates, $x$ and $y$, 
are the radius, $r$, and the 
$r \varphi$--coordinate at radius $r$,
which we denote as $u$.
The relation between these pairs of parameters reads:
\begin{equation}
\begin{array}{l}
x = r \cos \frac{\displaystyle u}
{\displaystyle r},~~~~~~~~~~~~~~~~~~~~~~~~~~~~~~~~~~~~~~~~~~~~~~~~~~~~~~\\
y = r \sin \frac{\displaystyle u}{\displaystyle r},
\end{array} 
\label{xyru}     
\end{equation} 
and 
\begin{equation}
\begin{array}{ll}
r & = \sqrt {x^2 + y^2}, \\
u & = r  \arctan \frac{\displaystyle y}{\displaystyle x}
= 2 r \arctan \frac{\displaystyle y}{\displaystyle r + x}
= 2 r \arctan \frac{\displaystyle r - x}{\displaystyle y}.  
\end{array} 
\label{ruxy}     
\end{equation}  
With the usage of an arc-length in the $xy$-plane, $s_t$, corresponding 
curvature $\omega$ and parameter $\lambda = \cot \theta$
(cotangent of the polar angle of the particle direction)  
\begin{equation}
\begin{array}{lll}
s_t = s \cdot n_t, & \omega = \frac{\displaystyle \omega_z}
{\displaystyle n_t}, &
 \lambda = \frac{\displaystyle n_z} {\displaystyle n_t},~~~~~~~~~~~~~~~~~~~~~~~~~~~~~~~
\end{array} 
\label{st}     
\end{equation} 
we obtain the solution for particle coordinates:
%\begin{equation}
%\begin{array}{l}
%\displaystyle
%x(s_t) = x_0 - \frac{1}{\omega} 
%\sin (\phi_0 - \omega s_t) + \frac{1}{\omega} \sin \phi_0 \\
%\displaystyle y(s_t) = y_0 + \frac{1}{\omega} 
%\cos (\phi_0 - \omega s_t) - \frac{1}{\omega} \cos \phi_0 \\
%z(s_t) = z_0 + \lambda_0 \cdot s_t.
%\end{array} 
%\label{xyzst}     
%\end{equation} 
%Finally, the solution of (\ref{xyzst}) takes the form: 
\begin{equation}
\begin{array}{l}\displaystyle
x(t) = r_0 \cos \frac{u_0}{r_0} - \frac{1}{\omega} 
\sin (\phi_0 - t)
+ \frac{1}{\omega} \sin \phi_0,~~~~~~~~~~~~~~~ \\
\displaystyle
y(t) = r_0 \sin \frac{u_0}{r_0} + \frac{1}{\omega} 
\cos (\phi_0 - t) 
- \frac{1}{\omega} \cos \phi_0, \\
\displaystyle
z(t) = z_0 + \frac{\lambda_0}{\omega} t,\\
t    = w \cdot s_t,\\
\end{array} 
\label{xyzt}     
\end{equation} 
where $r_0, u_0$ are values at $t=0$.
The particle which is located at a radius, $r_0$, given $t=0$, then 
arrives at a radius, $r$, given the value of $t$, which satisfies
the equation:
\begin{equation}
\begin{array}{ll}
r^2 = & r_0^2 + T + S\sin \alpha 
- (S \sin \alpha + T)\,\cos t + S \cos \alpha\,\sin t,\\
  & T = \frac{\displaystyle 2}{\displaystyle \omega^2},\,\,\,\, 
S = \frac{\displaystyle 2r_0}{\displaystyle \omega},\,\,\,\,
\alpha = \phi_0 - \frac{\displaystyle u_0}{\displaystyle r_0}.
\end{array} 
\label{r2}     
\end{equation} 
%page15
Solutions of the latter equation are
%time of arrivale at radius $r$ is equal:
\begin{equation}
\begin{array}{l}
t_{1,2} = 2 \arctan \left[ \frac{\displaystyle S \cos \alpha}
{\displaystyle D - 2\,T - 2\,S\,\sin \alpha} \left( 1 \pm
\sqrt{1 - 
\frac {\displaystyle D\cdot(D - 2\,T - 2\,S\,\sin \alpha)} 
{\displaystyle S^2 \cos^2 \alpha} }  \right)  \right] \\
D = r^2 - r_0^2 = \Delta r (2 r_0 + \Delta r),\,\,\,
\Delta r = r - r_0.
\end{array} 
\label{t}     
\end{equation} 
The solution $t_2$ (with minus sign) corresponds to a shorter path length.
%and will be used in the following. 
We describe a particle in a homogeneous magnetic field 
by a state vector at a reference cylindrical surface of radius $r_k$:
\begin{equation}
\begin{array}{l}
x^T_k = (u_k,\; z_k,\; \phi_k,\; \lambda_k,\; q_k)\,, 
\end{array} 
\label{xsol}     
\end{equation} 
where
\vspace*{-4mm}
\begin{itemize}
%\begin{enumerate}
\item[]\,\,$u_k$~~~=~~~ $r \varphi$--coordinate at radius $r_k$, 
\vspace*{-4mm}
\item[]\,\,$z_k$~~~=~~~$z$-coordinate,
\vspace*{-4mm}
\item[]\,\,$\phi_k$~~~=~~~angle of $xy$-projection of track direction with
  the $x$--axis,
\vspace*{-4mm}
\item[]\,\,$\lambda_k$~~~=~~~$\cot \theta$ at radius $r_k$, 
\vspace*{-4mm}
\item[]\,\,$q_k$~~~=~~~$Q/p_k$, inverse momentum signed according to particle 
charge, $Q$.
%\end{enumerate}
\end{itemize}
Such cylindrical parameterization looks natural for the
barrel tracking detectors.
An analogous state vector was used for 
the implementation
of the Kalman filter formalism for the 
ALEPH Time Projection Chamber~\cite{tlohse}.

%======================= 
 \subsection{Cylindrical Parameterization: Prediction Equations} 
%=======================
In the prediction stage of the Kalman filter,
the state vector ${x}_k$ is propagated at the next reference radius,
$r_{k+1} = r_k + \Delta r_k$. We obtain this transformation 
from (\ref{xyru}--\ref{xyzt}):
\begin{equation} 
\begin{array}{ll} 
u_{k+1} & = 2 r_{k+1} \arctan \frac{\displaystyle y_{k+1}}
{\displaystyle r_{k+1} + x_{k+1}} 
= r_{k+1}  \arctan \frac{\displaystyle y_{k+1}}{\displaystyle x_{k+1}}, \\ 
\displaystyle z_{k+1} & = z_k + \frac{\lambda_k}{\omega_k} t_k, \\
\phi_{k+1} & = \phi_{k} - t_k,\\
\lambda_{k+1} & = \lambda_{k}, \\
q_{k+1} & = q_{k},
\end{array}  
\label{xk1}      
\end{equation}   
where
\begin{equation} 
\begin{array}{ll}
x_{k+1} & = r_k \cos \displaystyle \frac{u_k}{r_k} - \frac{1}{\omega_k} 
\sin (\phi_k - t_k) 
+ \frac{1}{\omega_k} \sin \phi_k~~~~~~~~~~\\
y_{k+1} & = r_k \sin \displaystyle \frac{u_k}{r_k} + \frac{1}{\omega_k} 
\cos (\phi_k - t_k) 
- \frac{1}{\omega_k} \cos \phi_k, \\
\omega_k & = \kappa \cdot B_{zk} \cdot q_k \cdot \sqrt{1+\lambda_k^2}.\\
\end{array}  
\label{xyk1}      
\end{equation}   
and the variable, $t_k$, is evaluated from (\ref{t}).
%with
%\[
%T = \frac{\displaystyle 2}{\displaystyle \omega_k^2},\,\,\,\,  
%S = \frac{\displaystyle 2r_k}{\displaystyle \omega_k},\,\,\,\,
%D = \Delta r_k (2 r_k + \Delta r_k),\,\,\,\,
%\alpha = \phi_k - \frac{\displaystyle u_k}{\displaystyle r_k}. \]
We approximate 
the Jacobian of this transformation as:
\begin{equation}
\partial ({x}_{k+1})\,/\,\partial ({x_k})
= \left( \begin{array}{ccccc}
\partial u_{k+1}/\partial u_{k} &~~{\bf 0}~~&
\partial u_{k+1}/\partial \phi_{k} &
\partial u_{k+1}/\partial \lambda_{k} &
\partial u_{k+1}/\partial q_{k} \\
\partial z_{k+1}/\partial u_{k} &~~{\bf 1}~~&
\partial z_{k+1}/\partial \phi_{k} &
\partial z_{k+1}/\partial \lambda_{k} &
\partial z_{k+1}/\partial q_{k} \\
\partial \phi_{k+1}/\partial u_{k} &~~{\bf 0}~~&~~{\bf 1}~~&
\partial \phi_{k+1}/\partial \lambda_{k} & 
\partial \phi_{k+1}/\partial q_{k} \\ 
{\bf 0} & {\bf 0} & {\bf 0}  & {\bf 1} & {\bf 0} \\
{\bf 0} & {\bf 0} & {\bf 0}  & {\bf 0} & {\bf 1} \\
\end{array}
\right).
\label{jacob}
\end{equation}
%page16
Elements of the Jacobian which always are very close to zero or
unity, we set explicitly to 0 or 1, respectively. 
We exploit the sparse structure of the Jacobian to reduce
computations, as will be discussed in Sect.~\ref{fastkf}.
Nontrivial elements of the Jacobian are presented in appendix A.

%======================= 
 \subsection{Cylindrical Parameterization: Projection of State Vector
to MVD Measurement} 
%=======================
\label{prcylmvd}
The origin of the local coordinate system
of a MVD sensor is given by the vector $\vec{r}_c$.
The unit vector, $\vec{n}$, is perpendicular to the sensor plane.
We define the axis of measurement by the unit vector, $\vec{m}$, 
which is located in the sensor plane and is perpendicular to strips.
A state vector $x_k$ is defined at a cylindrical reference surface 
of a radius, $r_k$. We can define the radius, $r_k$, in such a way 
that the reference point will be close to the sensor. 
%where the measurement is recorded.     
In the immediate vicinity of the reference point,
we linearize equations (\ref{xk1},\ref{xyk1}) with respect to the
variable, $t_k$:
\begin{equation}
\begin{array}{ll}
x(t_k) & = x_k \displaystyle 
+ \frac{t_k}{\omega_k} \cos \phi_{k}, \\
 & \\
y(t_k) & = y_k \displaystyle
+ \frac{t_k}{\omega_k} \sin \phi_{k}, \\
& \\
z(t_k) & = 
\displaystyle
z_{k+1} + \frac{\lambda_{k}}{\omega_{k}}\,t_k. 
\end{array}  
\label{linxyz}      
\end{equation}   
A condition of the trajectory intersection with the sensor plane reads:
\begin{equation}
\begin{array}{l}
 \left[ \,( \vec{r}(t_k) - \vec{r}_c )  \cdot \vec{n} \,\right] = 0.
\end{array}  
\label{inters}      
\end{equation}   
The variable advance, $\Delta t_k$, to travel from the radius, $r_{k}$, to
the sensor plane is:
\begin{equation}
\begin{array}{ll}
\Delta t_k & = -\frac{\displaystyle b_k}{\displaystyle a_k}, \\
 & \\
a_k & = \frac{\displaystyle n_x}{\displaystyle \omega_k}\cos\phi_k
+ \frac{\displaystyle n_y}{\displaystyle \omega_k}\sin\phi_k
+ \frac{\displaystyle n_z}{\displaystyle \omega_k}\lambda_k, \\
 & \\
b_k & = (x_k - x_c)\,n_x + (y_k - x_c)\,n_y + (z_k - z_c)\,n_z.
\end{array}  
\label{dtk}      
\end{equation}   
To obtain the expected measurement, $h_k(x_k)$, we project the position vector
in the local frame, $\vec{r}(\Delta t_k) - \vec{r}_c$,
to the measurement axis, $\vec{m}$:
\begin{equation}
\begin{array}{ll}
h_k(x_k) 
 & =  \left[\,( \vec{r}(\Delta t_k) - \vec{r}_c ) \cdot\vec{m}\,\right]\\
 & = \frac{\displaystyle \Delta t_k}{\displaystyle \omega_k}\,c_k
+ (x_k - x_c)\,m_x + (y_k - y_c)\,m_y + (z_k - z_c)\,m_z, \\
&\\
~~~~~c_k  & = m_x\,\cos\phi_k + m_y\,\sin\phi_k + m_z\,\lambda_k.
\end{array}  
\label{h}      
\end{equation}
%page17   
Elements of the Jacobian, $\partial (h_k) / \partial (x_k)$, are:
\begin{equation}
\begin{array}{ll}
\partial h_k/ \partial u_k & = 
\frac{\displaystyle c_k}{\displaystyle \omega_k}
\,\partial\Delta t_k/\partial u_k 
- m_x \frac{\displaystyle y_k}{\displaystyle r_k}
+ m_y \frac{\displaystyle x_k}{\displaystyle r_k},\\ 
&\\
\partial h_k/ \partial z_k & =   
\frac{\displaystyle c_k\,n_z}{\displaystyle \omega_k\,a_k} + m_z,\\ 
&\\
\partial h_k/ \partial \phi_k & =  
\frac{\displaystyle c_k}{\displaystyle \omega_k} 
\,\partial\Delta t_k/\partial \phi_k 
+ \frac{\displaystyle \Delta t_k}{\displaystyle \omega_k}
( -m_x\sin\phi_k + m_y\cos\phi_k),\\ 
&\\
\partial h_k/ \partial \lambda_k & =   
\frac{\displaystyle \Delta t_k}{\displaystyle \omega_k}\,\left(
-\frac{\displaystyle n_z\,c_k}{\displaystyle a_k\,\omega_k} + m_z \right),\\ 
&\\
\partial h_k/ \partial q_k & = 0  ,\\ 
\end{array}  
\label{dhdx}      
\end{equation}   
with derivatives of $\Delta t_k$
\begin{equation}
\begin{array}{ll}
\partial \Delta t_k/ \partial u_k & =   
\frac{\displaystyle 1}{\displaystyle a_k}\,\left(
n_x\frac{\displaystyle y_k}{\displaystyle r_k}
- n_y \frac{\displaystyle x_k}{\displaystyle r_k} \right),
~~~~~~~~~~~~~~~~~~~~~~~~~~~~~~~~~~~\\   
&\\
\partial \Delta t_k/ \partial \phi_k & =   
\frac{\displaystyle \Delta t_k}{\displaystyle a_k\,\omega_k}\,\left(
n_x\,\sin\phi_k - n_y\,\cos\phi_k \right).\\    
\end{array}  
\label{dtdxhcyl}      
\end{equation}   
To exploit the sparse structure of the Jacobian  
and reduce computations 
we approximate the Jacobian  
%(\ref{dhdx}) of general
%structure $(\partial (h) / \partial (x))^T = ( h_1, h_2, h_3, h_4, {\bf 0})$ 
for specific cases:  
\begin{equation}
\begin{array}{l} 
\hder{(h_k)}{(x_k)} = \left(
\begin{array}{ccccc}
\dvder{h_k}{u_k}&~{\bf 1}&\dvder{h_k}{\phi_k}&\dvder{h_k}{\lambda_k}&~{\bf 0}\\
\end{array}
\right),~~{\rm for}~~m_z \approx 1,\\
\\
\hder{(h_k)}{(x_k)} = \left(
\begin{array}{ccccc}
\dvder{h_k}{u_k}&~{\bf 0}&\dvder{h_k}{\phi_k}&~~{\bf 0}&~~~{\bf 0}\\
\end{array}
\right),~~{\rm for}~~m_z \approx 0.\\
\end{array}  
\label{dhdxcylmvd}      
\end{equation}   

%======================= 
 \subsection{Cylindrical Parameterization: Projection of State Vector
to CTD Measurement} 
%=======================
\label{projcylctd}
Each sense stereo wire runs at a small angle, $\alpha$,
and its location in the $xy$-plane at coordinate $z$ is:
\begin{equation}
\begin{array}{ll}
\vec{w} & = \vec{r}_w + (z - z_c)\,\vec{r'}_w,\\
\end{array} 
\label{rsw}     
\end{equation} 
where $z_c$ is the $z$--coordinate of the nominal center of the CTD .
A ``planar drift'' approximation is used to render measurements 
in space~\cite{vctrack}. Drift distance is measured along the
``planar drift measurement axis'', $\vec{m}$: 
\begin{equation}
\begin{array}{ll}
m_x & = -n_{y} / | \vec{n} |, \\
m_y & = +n_{x} / | \vec{n} |, \\
\end{array} 
\label{mwxy}     
\end{equation} 
which is obtained by rotating the vector, $\vec{n}$, through $+90^\circ$.
The vector $\vec{n}$ depends linearly on the $z$ coordinate:
\begin{equation}
\begin{array}{ll}
\vec{n} & = \vec{p}_w + (z - z_c)\,\vec{p'}_w.\\
\end{array} 
\label{nsw}     
\end{equation} 
A state vector $x_k$ is defined at a cylindrical reference surface 
of a radius, $r_k$. We define the radius, $r_k$, in a way 
that the reference point is close to the point
where the trajectory 
%page18
hits the planar drift plane.   
Close to the reference point, 
we use linearized equations of motion (\ref{linxyz}).
A condition of the trajectory intersection with the planar drift plane reads:
\begin{equation}
\begin{array}{l}
 \left[ \,( \vec{r}(t_k) - \vec{w} )  \cdot \vec{n} \,\right] = 0.\\
\end{array}  
\label{intersctd}      
\end{equation}   
The variable advance, $\Delta t_k$, to travel from the radius, $r_{k}$, to
the planar drift plane is a solution (of smallest absolute value) 
of a quadratic equation,
$(\Delta t_k)^2\,a_k + \Delta t_k\,b_k + c_k = 0$:      
\begin{equation}
\begin{array}{l}
\Delta t_{k1,2} = \frac{\displaystyle 1}{\displaystyle 2 a_k}
\left( -b_k \pm \sqrt{b_k^2 - 4\,a_k\,c_k} \right),
\end{array}  
\label{quadrctd}      
\end{equation}   
with coefficients
\begin{equation}
\begin{array}{ll}
a_k = A_{kx}\,p'_{wx} + A_{ky}\,p'_{wy}, & \\ 
b_k = A_{kx}\,{\cal P}_{kx} + B_{kx}\,p'_{wx}
+ A_{ky}\,{\cal P}_{ky} + B_{ky}\,p'_{wy}, & \\
c_k = B_{kx}\,{\cal P}_{kx} + B_{ky}\,{\cal P}_{ky}, &  \\
A_{kx} = (\cos \phi_k - \lambda_k r'_{wx}) / \omega_k , &
A_{ky} = (\sin \phi_k - \lambda_k r'_{wy}) / \omega_k , \\
B_{kx} = x_k - r_{wx} - (z_k - z_c) r'_{wx}, & 
B_{ky} = y_k - r_{wy} - (z_k - z_c) r'_{wy}, \\
{\cal P}_{kx} = 
\left[ p_{wx} + (z_k - z_c) p'_{wx} \right] \omega_k / \lambda_k,&
{\cal P}_{ky} = 
\left[ p_{wy} + (z_k - z_c) p'_{wy} \right] \omega_k / \lambda_k. \\
\end{array}  
\label{coeffctd}      
\end{equation} 
The expected measurement, $h_k(x_k)$, is the drift distance.  
To evaluate it,
we project the position vector
in the planar drift system of the wire, $\vec{r}(\Delta t_k) - \vec{w}$,
to the measurement axis $\vec{m}$:
\begin{equation}
\begin{array}{l}
h_k(x_k) = 
 \left[ \,( \vec{r}(\Delta t_k) - \vec{w} )  \cdot \vec{m} \,\right].\\
\end{array}  
\label{projctd}      
\end{equation}   
To ``stretch'' the projected value  
according to the stereo angle, $\alpha$, we have to replace $\vec{m}$ by 
$\vec{m}/\cos\alpha$ in the following formulas.
%the stretched value of $\vec{m}$:
%\begin{equation}
%\begin{array}{l}        
%\vec{m'} = \vec{m} / \cos\alpha.
%\end{array}  
%\label{stratchcdt}      
%\end{equation}   
The expected measurement is a linear function of the $\Delta t_k$:
\begin{equation}
\begin{array}{ll} 
h_k(x_k)& =\, \Delta t_k\,{\cal C}_k + m_x B_{kx} + m_y B_{ky},\\
~~~~~{\cal C}_k& =\, (m_x A_{kx} + m_y A_{ky}) / \omega_k.
\end{array}  
\label{hkcylctd}      
\end{equation}   
We approximate the Jacobian, $\hder{(h_k)}{(x_k)}$,
by setting its elements which are very close to zero or
unity, explicitly to 0 or 1: 
\begin{equation}
\begin{array}{l} 
\hder{(h_k)}{(x_k)} = \left(
\begin{array}{ccccc}
{\bf 1}&\dvder{h_k}{z_k}&\dvder{h_k}{\phi_k}&\dvder{h_k}{\lambda_k}
&~{\bf 0} \\
\end{array}
\right),~~~~{\rm for}~~|\Delta t_k| \ge 10^{-6},\\
\\
\hder{(h_k)}{(x_k)} = \left(
\begin{array}{ccccc}
{\bf 1}&\dvder{h_k}{z_k}&~~~{\bf 0}&~~~{\bf 0}&~~~{\bf 0}\\
\end{array}
\right),~~~~{\rm for}~~|\Delta t_k|  < 10^{-6}.\\
\end{array}  
\label{dhdxcylctd}      
\end{equation}   
Nontrivial elements of the Jacobian are defined in appendix B.

The axial wires of the CTD run parallel to the $z$--axis and
parameters $\vec{r'}_w$ and $\vec{p'}_w$ vanish
in (\ref{rsw}) and (\ref{nsw}), respectively.
%Close to the reference point,  
%we use linearized equations of motion (\ref{linxyz}).
A condition of the intersection of the trajectory with
the ``planar drift plane'' results in Eq.~\ref{intersctd},
which has the solution  
\begin{equation}
\begin{array}{ll} 
\Delta t_k& = -b_k / a_k,\\
~a_k& = (\cos\phi_k\,p_{wx} + \sin\phi_k\,p_{wy}) / \omega_k,\\
~b_k& = (x_k - r_{wx})\,p_{wx} + (y_k - r_{wy})\,p_{wy}.\\
\end{array}   
\label{dtkcylax}       
\end{equation}    
%page19
A measurement vector for an axial wire, 
%in the CTD
$m_k$, is either one-dimensional
(drift distance) or two-dimensional (drift distance and z position).   
Let's consider the vector of expected measurement, $h_k(x_k)$, for
a general, two-dimensional case
\begin{equation}
h_k(x_k) = \left(
\begin{array}{l} 
h_{k1}(x_k)\\
h_{k2}(x_k)\\ 
\end{array}   
\right),
\label{hkcylaxial}       
\end{equation}    
with the first component (drift distance) and second (z position),
which are defined in (\ref{projctd})
and (\ref{linxyz}), respectively:
\begin{equation}
\begin{array}{l} 
h_{k1}(x_k) = 
(x_k + 
\frac{\displaystyle \Delta t_k}{\displaystyle \omega_k} \cos\phi_k 
- r_{wx})\,m_{wx} +
(y_k + 
\frac{\displaystyle \Delta t_k}{\displaystyle \omega_k} \sin\phi_k  
- r_{wy})\,m_{wy}, \\ 
h_{k2}(x_k) = z_k + \frac{\displaystyle \lambda_k}{\displaystyle \omega_k}\,
\Delta t_k.\\
\end{array}
\label{dzcylaxial}        
\end{equation}     
We approximate the Jacobian, $\partial(h_k)/\partial(x_k)$ as:
%set its elements, which are very close to zero or unity, 
%explicitly to 0 or 1: 
\begin{equation}
\begin{array}{l}
\partial (h_k) / \partial (x_k) = \left(
\begin{array}{ccccc}
~{\bf 1}&~{\bf 0}&\dvder{h_{k1}}{\phi_k}&{\bf 0}&~{\bf 0}\\
\dvder{h_{k2}}{u_k}&~{\bf 1}&\dvder{h_{k2}}{\phi_k}&
\dvder{h_{k2}}{\lambda_k}&~{\bf 0}\\
\end{array}
\right),~~{\rm for}~~|\Delta t_k| \ge 10^{-6}, \\
\end{array}
\label{jacobhcylaxial}      
\end{equation}
and 
\begin{equation}
\begin{array}{l}
\partial (h_k) / \partial (x_k) = \left(
\begin{array}{ccccc}
~{\bf 1}&~{\bf 0}&~~~{\bf 0}&~~~~{\bf 0}&~~~~{\bf 0}\\
\dvder{h_{k2}}{u_k}&~{\bf 1}&~~~{\bf 0}&~~~~{\bf 0}&~~~~{\bf 0}\\
\end{array}
\right),~~{\rm for}~~|\Delta t_k| < 10^{-6}. \\
\end{array}
\label{jacobhcylaxial1}      
\end{equation}
%We evaluate elements of the Jacobian in appendix B.
Elements of the Jacobian are presented in appendix B.

%======================= 
 \subsection{Cylindrical Parameterization: Process Noise} 
%=======================
We evaluate the components of a vector of particle direction, $\vec{n}$,  
using parameters $\phi,\,\lambda$:
\begin{equation} 
\begin{array}{l}
n_x = \frac{\displaystyle \cos \phi}
{\displaystyle \sqrt{1+\lambda^2}},~~~~
n_y = \frac{\displaystyle \sin \phi}
{\displaystyle \sqrt{1+\lambda^2}},~~~~
n_z = \frac{\displaystyle \lambda}
{\displaystyle \sqrt{1+\lambda^2}}~~~{\rm and}~~~
n_t = \frac{\displaystyle 1} 
{\displaystyle \sqrt{1+\lambda^2}}.
\end{array}  
\label{nphi}      
\end{equation}   
We obtain deviations of parameters $\phi,\,\lambda$,
induced by multiple scattering, from Eq.~(\ref{varn}):
\begin{equation}
\begin{array}{ll}
\delta \phi = \theta_1\,\sqrt{1+\lambda^2},&
~~~~\delta \lambda = -\theta_2\,\sqrt{1+\lambda^2},\\
\end{array}  
\label{varphi}      
\end{equation}   
where $\theta_{1}, \theta_{2}$ are random variables defined by (\ref{thetams}).
Nonzero elements of the matrix, describing
multiple scattering in one scatterer, are: 
\begin{equation}
\begin{array}{ll}
Q_{\phi\phi} = \theta_{ms}^2\,(1 + \lambda^2),&
~~~~Q_{\lambda\lambda} = \theta_{ms}^2\,(1 + \lambda^2),\\
\end{array}  
\label{Qij}      
\end{equation}   
with RMS of the deflection angle, $\theta_{ms}$, which is defined
by Eq.~(\ref{sms}).
The matrix, $Q_k$, in Eq.(\ref{ck1}) takes into account a summary effect 
of multiple scattering:
\begin{equation}
\begin{array} {l}
Q_k = {\displaystyle \sum_i F_{ik}\,Q_i\,F^T_{ik}},
~~~~{\rm with}~~~~F{ik} = \partial (x_k)/\partial(x_i),\\ 
\end{array}  
\label{Qijsum}   
\end{equation}   
and, therefore the index $i$ runs over all elements of material   
on the path from $(k-1)^{th}$ to $k^{th}$ state.  

%page20
%======================= 
 \section{Cartesian Parameterization in an Inhomogeneous Magnetic Field} 
% \section{Cartesian Parameterization for Forward Tracks} 
%=======================
\label{carfor}
The following choice of track parameters at a reference $z$--coordinate
is suited for forward tracks ($n_z > 0$):
\begin{equation}
\begin{array}{l}
\tilde{x}^T = (x,\; y,\; t_x,\; t_y,\; q),\\
\end{array}  
\label{xf}      
\end{equation}   
where
\vspace*{-4mm}
\begin{itemize}
%\begin{enumerate}
\item[]\,\,$x$~~~=~~~ $x$--coordinate in the Cartesian coordinate system 
of ZEUS, 
\vspace*{-4mm}
\item[]\,\,$y$~~~=~~~ $y$--coordinate in the Cartesian coordinate system,
\vspace*{-4mm}
\item[]\,\,$t_x$~~~=~~~$n_{x}/n_{z}$ track slope in $xz$--plane,
\vspace*{-4mm}
\item[]\,\,$t_y$~~~=~~~$n_{y}/n_{z}$ track slope in $yz$--plane, 
\vspace*{-4mm}
\item[]\,\,$q$~~~=~~~$Q/|\vec{p}|$, inverse momentum signed according 
to particle charge, $Q$.
%\end{enumerate}
\end{itemize}
This parametrization will be called ``cartesian''. 
The implementation of the Kalman filter technique in an inhomogeneous magnetic field
is analogous to those described in~\cite{tracing}.
In the following we discuss the case
of forward tracks. The rear tracks are specified in Subsec.~\ref{rear}.
%======================= 
\subsection{Cartesian Parametrization: Equations of Motion \\ 
in Inhomogeneous Magnetic Field} 
%=======================
\label{eqmotion}
For forward tracks we can use the $z$ coordinate as independent variable 
instead of the path length
%, $s$, %and replace $ds \to dz/n_z$.
in Eqs.~(\ref{dxds}).   
%The eqs.~(\ref{dxds}) 
The equations rewritten w.r.t. 
%the reference 
$z$ coordinate read:
\begin{equation}
\begin{array}{ll}
\label{dxdz}
dx/dz & = t_{x},  \\
dy/dz & = t_{y}, \\
dt_{x}/dz & = q \cdot \kappa \cdot A_{x}(t_{x},t_{y},\vec{B}),
~~~~~~~~~~~~~~~~~~~~~~~~~~~~~~ \\
dt_{y}/dz & = q \cdot \kappa \cdot A_{y}(t_{x},t_{y},\vec{B}), \\
q & = {\rm const}, 
\end{array}  
\end{equation}
where the functions $A_{x}$,$A_{y}$ are
\begin{equation}
\begin{array}{ll}
\begin{array}{l}
A_{x}=(1+t_{x}^{2}+t_{y}^2)^{\frac{1}{2}} \cdot 
\left[ t_{y}\cdot(t_{x}B_{x}+B_{z})
-(1+t_{x}^{2})B_{y} \right],\\
A_{y}=(1+t_{x}^{2}+t_{y}^2)^{\frac{1}{2}}\cdot 
\left[-t_{x}\cdot(t_{y}B_{y}+B_{z})
+(1+t_{y}^{2})B_{x} \right].\\
\end{array} 
\label{dxdz1}
\end{array}  
\end{equation}
To transport track parameters in the inhomogeneous field
from plane $z_{0}$ to plane $z$, we solve
%use Runge-Kutta method for numerical solving of 
the latter equations with initial values defined at $z_0$  
\begin{equation}
\label{x0}
\tilde{x}_{0}^T=(x_{0},\; y_{0},\; t_{x0},\; t_{y0},\; q_{0}). 
~~~~~~~~~~~~~~~~~~~~~~~~~~~~~~~~~~~~~~
\end{equation}
%Let us assume we should propagate the particle parameters from plane $z_{0}$
%to plane $z$.  
Three methods are used to solve Eqs. (\ref{dxdz}), 
depending on the distance, $s=z-z_{o},$ between these planes.  

1. {\em $\left|s\right| < 10~{\rm cm}$}: a parabolic expansion of the particle 
trajectory is used

%page21
\begin{equation}
%\[  
\begin{array}{cl}
x(z)= & x_{0}+t_{x0}\cdot s + \frac{1}{2} \cdot q_{0}\cdot \kappa 
\cdot A_{x} \cdot s^{2},\\    
y(z)= &y_{0}+t_{y0}\cdot s + \frac{1}{2} \cdot q_{0} 
\cdot \kappa \cdot A_{y} \cdot s^{2},\\
t_{x}(z)= & t_{x0} + q_{0} \cdot \kappa \cdot A_{x} \cdot s,\\
t_{y}(z)= & t_{y0} + q_{0} \cdot \kappa \cdot A_{y} \cdot s,\\ 
q(z)= & q_{0}.\\ 
\end{array}    
%\]
\label{parabolic}
\end{equation}

2. {\em$ 10~{\rm cm} \leq \left|s\right| < 60~{\rm cm}$}: 
the classical fourth-order 
Runge-Kutta method ~\cite{RK} is selected to find the solution of the
equations (\ref{dxdz}) .

3. {\em $\left|s\right| \geq 60~{\rm cm}$}: a fifth-order Runge-Kutta method
with adaptive step size control ~\cite{RK} is used.

%======================= 
 \subsection{Cartesian Parametrization: Equations for Derivatives} 
%=======================
The Jacobian of transformation of parameters given at $z_0$ to $z$,
$\hder{(\tilde{x})}{(\tilde{x}_{0})}$, is defined as:
\begin{equation}
\hder{(\tilde{x})}{(\tilde{x}_{0})}
= \left( \begin{array}{ccccc}
~~{\bf 1}~~ & ~~{\bf 0}~~& ~\dvder{x}{t_{x0}}~ & ~\dvder{x}{t_{y0}}~ & ~\dvder{x}{q_0}~ \\
{\bf 0} & {\bf 1} & \dvder{y}{t_{x0}} & \dvder{y}{t_{y0}} & \dvder{y}{q_0} \\
{\bf 0} & {\bf 0} & {\bf 1} & \dvder{t_x}{t_{y0}} & \dvder{t_x}{q_0} \\
{\bf 0} & {\bf 0} & \dvder{t_y}{t_{x0}} & {\bf 1} & \dvder{t_y}{q_0} \\
{\bf 0} & {\bf 0} & {\bf 0} & {\bf 0} & {\bf 1}, \\
\end{array}
\right).
\label{jacobcar}
\end{equation}
Elements of the latter Jacobian which 
%always 
are very close to zero or
unity, are set to 0 or 1, respectively. 
Nontrivial elements of the Jacobian (\ref{jacobcar}) 
for short distance 
%using the parabolic expansion 
$(\left|s\right| < 10~{\rm cm})$  
we approximate as:
\begin{equation}
\begin{array}{ll}
\hder{x}{t_{x0}} = s, &
~~~\hder{x}{t_{y0}} = \dvrat{1}{2}\,q_0\,\kappa\,s^2\,\dvder{A_x}{t_{y0}},\\
\hder{y}{t_{x0}} = \dvrat{1}{2}\,q_0\,\kappa\,s^2\,\dvder{A_y}{t_{x0}},&
~~~\hder{y}{t_{y0}} = s, \\
\hder{t_x}{t_{y0}} = q_0\,\kappa\,s\,\dvder{A_x}{t_{y0}},&
~~~\hder{t_y}{t_{x0}} = q_0\,\kappa\,s\,\dvder{A_y}{t_{x0}},\\
\hder{x}{q_0} = \dvrat{1}{2}\,\kappa\,s^2\,A_x, & 
~~~\hder{y}{q_0} = \dvrat{1}{2}\,\kappa\,s^2\,A_y,\\ 
\hder{t_x}{q_0} = \kappa\,s\,A_x, &
~~~\hder{t_y}{q_0} = \kappa\,s\,A_y,\\
\end{array}
\label{jacobparcar}
\end{equation}
with derivatives $\hder{A_x}{t_{y0}}$ and $\hder{A_y}{t_{x0}}$,
which we define below. 

To swim derivatives at long distance ($\left|s\right| \geq 10\,{\rm cm}$),
we define equations for derivatives as described in \cite{tracing} 
and solve them
by a Runge-Kutta method simultaneously with 
%Eqs.\,(\ref{dxdz}).
equations of motion.
The magnetic field is smooth enough even in the STT area
and, therefore we regard Eqs.\,(\ref{dxdz}) as
almost invariant with respect to small shifts by $x$ and $y$. 
%This means that
Derivatives with respect to initial $x_{0}$, $y_{0}$ are trivial :
\[  \begin{array}{cl}
\partial \tilde{x}^T / \partial x_{0}= &
( 1,\; 0,\; 0,\; 0,\; 0 ), \\
\partial \tilde{x}^T / \partial y_{0}= &
( 0,\; 1,\; 0,\; 0,\; 0 ). 
\end{array}    \]
%page22
To obtain equations for $\partial \tilde{x} / \partial t_{x0}$, we
differentiate  equations (\ref{dxdz}) with respect to $t_{x0}$ and change
the order of the derivative operators $\partial  / \partial t_{x0}$ 
and $d/dz$ on the left hand sides :
\begin{equation}
\begin{array}{ll}
\label{dtx}
d/dz(\partial x / \partial t_{x0}) = & \partial t_{x}/\partial t_{x0},  \\
d/dz(\partial y / \partial t_{x0}) = & \partial t_{y}/\partial t_{x0}, \\
d/dz(\partial t_{x} / \partial t_{x0}) = & 
 q_{0} \cdot \kappa \cdot \left[(\partial A_{x}/\partial t_{x})
( \partial t_{x}/\partial t_{x0}) 
+(\partial A_{x}/\partial t_{y})
( \partial t_{y}/\partial t_{x0}) \right], \\

d/dz(\partial t_{y} / \partial t_{x0}) = & 
 q_{0} \cdot \kappa \cdot \left[(\partial A_{y}/\partial t_{x})
( \partial t_{x}/\partial t_{x0}) 
+(\partial A_{y}/\partial t_{y})
( \partial t_{y}/\partial t_{x0}) 
\right], \\
\partial q / \partial t_{x0} \ = & 0,
\end{array}  
\end{equation}
where
\[  \begin{array}{ll}
\partial A_{x}/\partial t_{x}= t_{x} \cdot A_{x}/(1+t_{x}^{2}+t_{y}^{2})
+(1+t_{x}^{2}+t_{y}^{2})^{\frac{1}{2}} \cdot
(t_{y} \cdot B_{x} - 2\cdot t_{x}\cdot B_{y}) \; , \\
\partial A_{x}/\partial t_{y}= t_{y} \cdot A_{x}/(1+t_{x}^{2}+t_{y}^{2})
+(1+t_{x}^{2}+t_{y}^{2})^{\frac{1}{2}} \cdot
(t_{x} \cdot B_{x} + B_{z}) \; , \\
\partial A_{y}/\partial t_{x}= t_{x} \cdot A_{y}/(1+t_{x}^{2}+t_{y}^{2})
+(1+t_{x}^{2}+t_{y}^{2})^{\frac{1}{2}} \cdot
(-t_{y} \cdot B_{y} - B_{z}) \; , \\
\partial A_{y}/\partial t_{y}= t_{y} \cdot A_{y}/(1+t_{x}^{2}+t_{y}^{2})
+(1+t_{x}^{2}+t_{y}^{2})^{\frac{1}{2}} \cdot
(-t_{x} \cdot B_{y} + 2\cdot t_{y}\cdot B_{x}) \; . \\
\end{array}  
 \;\;\;\;\;\; 
%(4)   
\]
Initial values for the solution of latter equations are:
\begin{equation}
\label{dtx0}
\partial \tilde{x}^T / \partial t_{x0} = 
( 0,\; 0,\; 1,\; 0,\; 0 ).
\end{equation}
The equations for $\partial \tilde{x} / \partial t_{y0}$ are analogous
to Eqs.\,(\ref{dtx}) , but the initial values  are :
\[  \begin{array}{cl}
\partial \tilde{x}^T / \partial t_{y0}= &
( 0,\; 0,\; 0,\; 1,\; 0 ) \;\;. \;\;\;\;\;
%(6)
\\
\end{array}    \]
To obtain equations for $\partial \tilde{x} / \partial q_{0}$, we
differentiate  Eqs.\,(\ref{dxdz}) with respect to $q_{0}$ and change
the order of the derivative operators 
$\partial  / \partial q_{0}$ and $d/dz$
in the left parts : 
\begin{equation}
\begin{array}{ll}
\label{dq}
d/dz(\partial x / \partial q_{0}) = & \partial t_{x}/\partial q_{0} \;,  \\
d/dz(\partial y / \partial q_{0}) = & \partial t_{y}/\partial q_{0} \;, \\
d/dz(\partial t_{x} / \partial q_{0}) = & 
  \kappa \cdot A_{x} +  \kappa \cdot q_{0} \cdot
\left[(\partial A_{x}/\partial t_{x})
( \partial t_{x}/\partial q_{0}) 
+(\partial A_{x}/\partial t_{y})
( \partial t_{y}/\partial q_{0}) 
\right]  \;, \\

d/dz(\partial t_{y} / \partial q_{0}) = & 
 \kappa \cdot A_{y} + \kappa \cdot q_{0} \cdot
\left[(\partial A_{y}/\partial t_{x})
( \partial t_{x}/\partial q_{0}) 
+(\partial A_{y}/\partial t_{y})
( \partial t_{y}/\partial q_{0}) 
\right]  \;, \\
\partial q / \partial q_{0} = & 1  \; .
\end{array}  
\end{equation}
Initial values for the solution of latter equations are :
\begin{equation}
\label{dq0}
\partial \tilde{x}^T / \partial q_{0} = 
( 0,\; 0,\; 0,\; 0,\; 1 ).
\end{equation}

%======================= 
 \subsection{Cartesian Parameterization: Projection of State Vector
to MVD Measurement} 
%=======================
\label{projcarmvd}
To project a state vector (\ref{xf})
to a BMVD measurement we use the method described in
Subsect.\,\ref{prcylmvd}. The state vector, $\tilde{x}_k$,
is defined in the reference plane with coordinate, $z=z_k$. 
We locate the reference plane 
%in a way, that the reference point is 
close to 
%the plane of 
the MVD sensor
and, therefore use a linear expansion of the trajectory:  
\begin{equation}
\begin{array}{ll}
x(s_k) & = x_k + t_{xk}\,s_k\\
y(s_k) & = y_k + t_{yk}\,s_k\\
z(s_k) & = z_k + s_k. 
\end{array}  
\label{linxyzcar}      
\end{equation}
%page23   
A condition of the trajectory intersection with the sensor plane reads:
\begin{equation}
\begin{array}{l}
 \left[ \,( \vec{r}(s_k) - \vec{r}_c )  \cdot \vec{n} \,\right] = 0,
\end{array}  
\label{interscar}      
\end{equation}   
where $\vec{r}_c$ and $\vec{n}$ are the origin of a local MVD sensor system and
the unit vector which is perpendicular to the sensor plane, respectively. 
The variable advance, $\Delta s_k$, to travel from the 
reference plane at $z_k$ to the sensor plane is:
\begin{equation}
\begin{array}{ll}
\Delta s_k & = -\frac{\displaystyle b_k}{\displaystyle a_k}, \\
~~~a_k & = t_{xk}\,n_x + t_{yk}\,n_y + n_z,\\
~~~b_k & = (x_k - x_c)\,n_x + (y_k - x_c)\,n_y + (z_k - z_c)\,n_z.
\end{array}  
\label{dsk}      
\end{equation}   
Analogous to Eq.\,\ref{h},  
we obtain the expected measurement, $h_k(\tilde{x}_k)$,
by projecting the position vector
in the local frame, $\vec{r}(\Delta s_k) - \vec{r}_c$,
to the measurement axis, $\vec{m}$:
\begin{equation}
\begin{array}{ll}
h_k(\tilde{x}_k) 
 & =  \left[\,( \vec{r}(\Delta s_k) - \vec{r}_c ) \cdot\vec{m}\,\right]\\
 & = \frac{\displaystyle \Delta s_k}{\displaystyle \omega_k}\,c_k
+ (x_k - x_c)\,m_x + (y_k - y_c)\,m_y + (z_k - z_c)\,m_z, \\
&\\
~~~~c_k  & = m_x\,t_{xk} + m_y\,t_{yk} + m_z.
\end{array}  
\label{hcar}      
\end{equation}   
Nontrivial elements of the Jacobian, 
$\partial (h_k) / \partial (\tilde{x}_k)$, are: 
\begin{equation} 
\begin{array}{ll} 
\hder{h_k}{x_k} = m_x - c_k\,n_x\,/\,a_k, & 
~~~~\hder{h_k}{y_k} = m_y - c_k\,n_y\,/\,a_k, \\ 
\hder{h_k}{t_{xk}} = \hder{h_k}{x_k}\cdot\Delta s_k, & 
~~~~\hder{h_k}{t_{yk}} = \hder{h_k}{y_k}\cdot\Delta s_k. \\ 
\end{array}   
\label{dhdxcar}       
\end{equation}    
Derivatives with respect to slopes 
%in Eqs.\,\ref{dhdxcar}
have an additional order of smallness $o(\Delta s_k)$ and 
we approximate the Jacobian, $\hder{(h_k)}{(\tilde{x}_k)}$,
for the BMVD:
\begin{equation}
\begin{array}{l} 
\hder{(h_k)}{(\tilde{x}_k)} = \left(
\begin{array}{ccccc}
\dvder{h_k}{x_k}&\dvder{h_k}{y_k}&\dvder{h_k}{t_{xk}}&
\dvder{h_k}{t_{yk}}&~{\bf 0} \\
\end{array}
\right),~~~~{\rm for}~~|\Delta s_k| \ge 10^{-3},\\
\\
\hder{(h_k)}{(\tilde{x}_k)} = \left(
\begin{array}{ccccc}
\dvder{h_k}{x_k}&\dvder{h_k}{y_k}&~~~{\bf 0}&~~~~{\bf 0}&~~~{\bf 0}\\
\end{array}
\right),~~~~{\rm for}~~|\Delta s_k|  < 10^{-3}.\\
\end{array}  
\label{dhdxcarmmvd}      
\end{equation}   
Sensors of the FMVD are almost perpendicular to the $z$--axis and, therefore
$n_{x,y} \approx 0$.
We locate the reference plane at the position of the 
FMVD sensor ($\Delta s_k=0$). Taking into account latter remarks,
we obtain from Eq.\,(\ref{dhdxcar}) 
the Jacobian, $\hder{(h_k)}{(\tilde{x}_k)}$, for the FMVD
\begin{equation}
\begin{array}{l} 
\hder{(h_k)}{(\tilde{x}_k)} = \left(
\begin{array}{ccccc}
~m_x~&~m_y~&~~{\bf 0}~~&~{\bf 0}~&~~{\bf 0}\\ 
\end{array}
\right).~~~~~~~~~~~~~~~~~~~~~~~~
\end{array}
\label{dhdxcarfmvd}      
\end{equation}   

%======================= 
 \subsection{Cartesian Parameterization: Projection of State Vector
to CTD Measurement} 
%=======================
\label{projcarctds}
The linear expansion of a particle trajectory (\ref{linxyzcar}) defines 
the particle coordinates in the immediate vicinity of a stereo wire.  
An approach to obtain the projection of cartesian 
%page24
state vector
to CTD stereo measurement is similar to those
discussed in Subsec.~\ref{projcylctd}. 
A condition of the trajectory intersection with the planar drift plane reads:
\begin{equation}
\begin{array}{l}
 \left[ \,( \vec{r}(s_k) - \vec{w} )  \cdot \vec{n} \,\right] = 0,\\
\end{array}  
\label{intersctdcar}      
\end{equation}   
where the coordinate of the wire, $\vec{w}$, 
and vector, $\vec{n}$, are defined by (\ref{rsw}) and (\ref{nsw}),
respectively.
The variable advance, $\Delta s_k$, to travel from the reference plane to
the planar drift plane is a solution 
of a quadratic equation, $(\Delta s_k)^2\,a_k + \Delta s_k\,b_k + c_k = 0$:
\begin{equation}
\begin{array}{l}
\Delta s_{k} = \frac{\displaystyle 1}{\displaystyle 2 a_k}
\left( -b_k + \sqrt{b_k^2 - 4\,a_k\,c_k} \right),
\end{array}  
\label{quadrctdcar}      
\end{equation}   
with coefficients
\begin{equation}
\begin{array}{ll}
a_k = A_{kx}\,p'_{wx} + A_{ky}\,p'_{wy}, & \\ 
b_k = A_{kx}\,{\cal P}_{kx} + B_{kx}\,p'_{wx}
+ A_{ky}\,{\cal P}_{ky} + B_{ky}\,p'_{wy}, & \\
c_k = B_{kx}\,{\cal P}_{kx} + B_{ky}\,{\cal P}_{ky}, &  \\
A_{kx} = t_{kx} - r'_{wx}, &
A_{ky} = t_{ky} - r'_{wy}, \\
B_{kx} = x_k - r_{wx} - (z_k - z_c)\,r'_{wx}, & 
B_{ky} = y_k - r_{wy} - (z_k - z_c)\,r'_{wy}, \\
{\cal P}_{kx} = 
p_{wx} + (z_k - z_c)\,p'_{wx},&
{\cal P}_{ky} = 
p_{wy} + (z_k - z_c)\,p'_{wy}. \\
\end{array}  
\label{coeffctdcar}      
\end{equation}   
We obtain 
the expected measurement, $h_k(x_k)$, 
i.e. drift distance, as in (\ref{projctd}):
\begin{equation}
\begin{array}{ll} 
h_k(x_k)& =~~\left[x_k + t_{kx}\,\Delta s_k - r_{wx} 
	- (z_k-z_c+\Delta s_k)\,r'_{wx} \right]\,m_x \\
&
~~+ \left[y_k + t_{ky}\,\Delta s_k - r_{wy}  
        - (z_k-z_c+\Delta s_k)\,r'_{wy} \right]\,m_y, \\
\end{array}  
\label{hkcylctdcar}      
\end{equation}
where the $\vec{m}$ have to by replaced by $\vec{m}/\cos\alpha$
to take into account the stereo angle, $\alpha$.
Derivatives with respect to slopes 
%in Eqs.\,\ref{dhdxcar}
have an additional order of smallness $o(\Delta s_k)$ and 
we approximate the Jacobian, $\hder{(h_k)}{(\tilde{x}_k)}$,
for the CTD stereo measurement:
\begin{equation}
\begin{array}{l} 
\hder{(h_k)}{(\tilde{x}_k)} = \left(
\begin{array}{ccccc}
\dvder{h_k}{x_k}&\dvder{h_k}{y_k}&\dvder{h_k}{t_{xk}}&
\dvder{h_k}{t_{yk}}&~{\bf 0} \\
\end{array}
\right),~~~~{\rm for}~~|\Delta s_k| \ge 10^{-3},\\
\\
\hder{(h_k)}{(\tilde{x}_k)} = \left(
\begin{array}{ccccc}
\dvder{h_k}{x_k}&\dvder{h_k}{y_k}&~~~{\bf 0}&~~~~{\bf 0}&~~~{\bf 0}\\
\end{array}
\right),~~~~{\rm for}~~|\Delta s_k|  < 10^{-3}.\\
\end{array}  
\label{hcarctds}      
\end{equation}   
Elements of the Jacobian, 
$\partial (h_k) / \partial (\tilde{x}_k)$, are presented in appendix C. 

Axial wires of the CTD run parallel to the $z$--axis and
parameters, $\vec{r'}_w$ and $\vec{p'}_w$ vanish
in (\ref{rsw}) and (\ref{nsw}), respectively.
A condition of the intersection of the trajectory with
the planar drift plane leads to Eq.~(\ref{intersctdcar}),
which has the solution:  
\begin{equation}
\begin{array}{ll} 
\Delta s_k& = -b_k / a_k,\\
~a_k& = t_{xk}\,p_{wx} + t_{yk}\,p_{wy},\\
~b_k& = (x_k - r_{wx})\,p_{wx} + (y_k - r_{wy})\,p_{wy}.\\
\end{array}   
\label{dtkcarax}       
\end{equation}    
We consider the vector of expected measurement, $h_k(x_k)$, for
the general, two-dimensional case
\begin{equation}
h_k(x_k) = \left(
\begin{array}{l} 
h_{k1}(x_k)\\
h_{k2}(x_k)\\ 
\end{array}   
\right),
\label{hkcaraxial}       
\end{equation} 
%page25   
with the first and second component
being a drift distance and $z$--position, respectively:
\begin{equation}
\begin{array}{l} 
h_{k1}(x_k) = 
\left(x_k + t_{kx}\,\Delta s_k - r_{wx} \right)\,m_x 
+ \left(y_k + t_{ky}\,\Delta s_k - r_{wy} \right)\,m_y, \\ 
h_{k2}(x_k) = z_k + \Delta s_k.\\
\end{array}
\label{dzcaraxial}        
\end{equation}     
We approximate the Jacobian, $\partial(h_k)/\partial(x_k)$, as: 
\begin{equation}
\begin{array}{l}
\partial (h_k) / \partial (x_k) = \left(
\begin{array}{ccccc}
\dvder{h_{k1}}{x_k}&\dvder{h_{k1}}{y_k}&\dvder{h_{k1}}{t_{kx}}
&\dvder{h_{k1}}{t_{ky}}&~{\bf 0}\\
\dvder{h_{k2}}{x_k}&\dvder{h_{k2}}{y_k}&\dvder{h_{k2}}{t_{kx}}&
\dvder{h_{k2}}{t_{ky}}&~{\bf 0}\\
\end{array}
\right),~~{\rm for}~~|\Delta s_k| \ge 10^{-3}, \\
\end{array}
\label{jacobcaraxial}      
\end{equation}
and 
\begin{equation}
\begin{array}{l}
\partial (h_k) / \partial (x_k) = \left(
\begin{array}{ccccc}
\dvder{h_{k1}}{x_k}&\dvder{h_{k2}}{y_k}&~~~{\bf 0}&~~~~{\bf 0}&~~~~{\bf 0}\\
\dvder{h_{k2}}{x_k}&\dvder{h_{k2}}{y_k}&~~~{\bf 0}&~~~~{\bf 0}&~~~~{\bf 0}\\
\end{array}
\right),~~{\rm for}~~|\Delta s_k| < 10^{-3}, \\
\end{array}
\label{jacobcaraxial1}      
\end{equation}
where we take into account an additional order of smallness $o(\Delta s_k)$
for derivatives with respect to track slopes.
We obtain nontrivial elements of latter Jacobians in appendix C. 

%======================= 
 \subsection{Cartesian Parameterization: Projection of State Vector
to STT Measurement} 
%=======================
Signal wires of a given STT layer are arranged in a plane 
perpendicular to the $z$--axis with coordinate $z=z_w$.
We locate the reference plane at the position of the layer, i.e.  
$z_k=z_w$.
The particle trajectory 
inside a straw tube we approximate by a straight line.
The latter line and the 
signal wire are described 
as lines which pass through points $\vec{r}_k$ and $\vec{r}_w$ 
and have directions $\vec{n}_k$ and $\vec{n}_w$, respectively:
\begin{equation}
\begin{array}{llll}
\vec{r}_k = \left(\begin{array}{c} x_k\\y_k\\z_w\\ \end{array}\right),&
\vec{r}_w = \left(\begin{array}{c} x_w\\y_w\\z_w\\ \end{array}\right),&
\vec{n}_k = \left(\begin{array}{c} n_{kx}\\n_{ky}\\n_{kz}\\ \end{array}
\right),&
\vec{n}_w = \left(\begin{array}{c} n_{wx}\\n_{wy}\\0\\ \end{array}\right).\\
\end{array}
\label{lineeq}      
\end{equation}   
Components of the vector of particle direction, $\vec{n}_k$, we calculate 
using track slopes $t_{kx},\,t_{ky}$:
\begin{equation} 
\begin{array}{l}
n_{kx} = \displaystyle \frac{t_{kx}}
	{\displaystyle \sqrt{1 + t_{kx}^2 + t_{ky}^2}},~~~
n_{ky} = \displaystyle \frac{t_{ky}}
	{\displaystyle \sqrt{1 + t_{kx}^2 + t_{ky}^2}},~~~
n_{kz} = \displaystyle \frac{1}
	{\displaystyle \sqrt{1 + t_{kx}^2 + t_{ky}^2}}.
\\
\end{array}  
\label{ntx}      
\end{equation}   
The expected measurement is a drift distance\footnote{
We expect that left--right ambiguity of the drift distance is resolved and, 
therefore regard it as a signed value.}
in the straw,
which is evaluated as a distance between these two lines:
\begin{equation}
\begin{array}{ll}
h_k(\tilde{x}_k) & =~~\displaystyle \frac
{\displaystyle (\vec{r}_k - \vec{r}_w) \cdot \vec{n}_k \times \vec{n}_w} 
{\displaystyle \left| \vec{n}_k \times \vec{n}_w \right|}. \\
\\
\end{array} 
\label{hsttcar}      
\end{equation}   
%page26
After simple calculations the expected measurement reads:
\begin{equation}
\begin{array}{ll}
h_k(\tilde{x}_k) & =~~\displaystyle \frac 
{\displaystyle -(x_k - x_w) n_{wy} + (y_k - y_w) n_{wx}}
{\displaystyle \sqrt{1 + {(t_{kx}\,n_{wy} - t_{ky}\,n_{wx})}^2}}.
\\
\end{array} 
\label{hsttcar1}      
\end{equation}   
The Jacobian of the latter transformation we can approximate as:
\begin{equation}
\begin{array}{l} 
\hder{(h_k)}{(\tilde{x}_k)} = \left(
\begin{array}{ccccc}
\dvder{h_k}{x_k}&\dvder{h_k}{y_k}&~~~{\bf 0}&~~~~{\bf 0}&~~~{\bf 0}\\
\end{array}
\right),\\
\end{array}  
\label{dhdxcarstt}      
\end{equation}   
with
\begin{equation} 
\begin{array}{l} 
\hder{h_k}{x_k} = \displaystyle -n_{wy}\,/\,
\displaystyle \sqrt{1 + {(t_{kx}\,n_{wy} - t_{ky}\,n_{wx})}^2}, \\
\\
\hder{h_k}{y_k} = ~~\displaystyle n_{wx}\,/\,
\displaystyle \sqrt{1 + {(t_{kx}\,n_{wy} - t_{ky}\,n_{wx})}^2}. \\
\end{array}   
\label{dhdxstt}       
\end{equation}    
%======================= 
 \subsection{Cartesian Parameterization: Process Noise} 
%=======================
We evaluate deviations of track slopes 
%$t_{kx},\,t_{ky}$
induced by multiple scattering from Eq.~(\ref{varn}):
\begin{equation}
\begin{array}{ll}
\delta t_{kx} = \displaystyle \delta \left(\dvrat{n_{kx}}{n_{kz}} \right) & 
=-\theta_1 \dvrat{n_{ky}}{n_{kz}\,n_{kt}} 
  + \theta_2 \dvrat{n_{kx}}{n_{kz}^2\,n_{kt}}, \\
\delta t_{ky} = \displaystyle \delta \left(\dvrat{n_{ky}}{n_{kz}} \right) & 
=~~~\theta_1 \dvrat{n_{kx}}{n_{kz}\,n_{kt}} 
  + \theta_2 \dvrat{n_{ky}}{n_{kz}^2\,n_{kt}}, \\
\end{array}  
\label{vartx}      
\end{equation}   
where $\theta_{1}, \theta_{2}$ are random variables defined by (\ref{thetams}).
Nonzero elements of the matrix describing
multiple scattering in one scatterer are: 
\begin{equation}
\begin{array}{l}
\displaystyle
Q_{t_x\,t_x} = \theta_{ms}^2\,(1 + t_{kx}^2)\,(1 + t_{kx}^2 + t_{ky}^2),
~~~~~~~~~~\\
\displaystyle
Q_{t_y\,t_y} = \theta_{ms}^2\,(1 + t_{ky}^2)\,(1 + t_{kx}^2 + t_{ky}^2),\\
\displaystyle
Q_{t_x\,t_y} = \theta_{ms}^2\,t_{kx}\,t_{ky}\,(1 + t_{kx}^2 + t_{ky}^2),\\
\end{array}  
\label{Qcarij}      
\end{equation}   
with RMS of the deflection angle, $\theta_{ms}$, which is defined
by Eq.~(\ref{sms}).
The matrix, $Q_k$, in prediction equation (\ref{ck1}) has to account for 
a summary effect of multiple scattering
on a path from $(k-1)^{th}$ to $k^{th}$ state,
and is therefore evaluated analogous to (\ref{Qijsum}).  

%======================= 
 \subsection{Cartesian Parameterization for Rear Tracks} 
%=======================
\label{rear}
For rear tracks $(n_z < 0)$ we use a parameterization 
analogous to those 
%described in Sect.~\ref{carfor} 
for forward tracks.
The meaning of parameters $x,\,y,\,q$ is identical with (\ref{xf}).
For rear tracks we define slopes w.r.t. negative direction of the $z$--axis:
\begin{equation} 
\begin{array}{l} 
t_x = - n_x\,/\,n_z,\\
t_y = - n_y\,/\,n_z.\\
\end{array}   
\label{xfrear}       
\end{equation}    
Equations of particle motion for rear tracks are
identical to (\ref{dxdz}) for forward tracks, but with slightly different
definition of functions $A_{x}$,$A_{y}$:
\begin{equation}
\begin{array}{ll}
\begin{array}{l}
A_{x}=(1+t_{x}^{2}+t_{y}^2)^{\frac{1}{2}} \cdot 
\left[ t_{y}\cdot(-t_{x}B_{x}+B_{z})
+(1+t_{x}^{2})B_{y} \right],\\
A_{y}=(1+t_{x}^{2}+t_{y}^2)^{\frac{1}{2}}\cdot 
\left[-t_{x}\cdot(-t_{y}B_{y}+B_{z})
-(1+t_{y}^{2})B_{x} \right].\\
\end{array} 
\label{dxdz1rear}
\end{array}  
\end{equation}
Equations (\ref{linxyzcar}) 
for linear and (\ref{parabolic}) for parabolic expansions of trajectory
can be used for rear tracks also, if we regard the expansion w.r.t.
$z$--coordinate decrement, $s=z_0-z$. 
%page27
%======================= 
 \section{Global Parameterization}  
%=======================
A global perigee parameterization of tracks~\cite{vctrack} is used for 
analyses in the ZEUS experiment.
The  perigee parameters are parameters of a helix, which are  
defined at the track's point of closest approach to the $z$--axis: 
\begin{equation}
\begin{array}{l}
\tilde{x}^T = (\phi_H,\; Q/R_H,\; QD_H,\; z_H,\; \cot \theta)\,, 
\end{array} 
\label{xsperigee}     
\end{equation} 
where
\vspace*{-4mm}
\begin{itemize}
%\begin{enumerate}
\item[]\,\,$\phi_H$~~~~~~=~~angle of $xy$--projection of track direction with
the $x$--axis,
\vspace*{-4mm}
\item[]\,\,$Q/R_H$~=~~helix curvature signed by a particle charge, $Q$,
\vspace*{-4mm}
\item[]\,\,$QD_H$~~=~~signed minimal distance to $z$--axis,
\vspace*{-4mm}
\item[]\,\,$z_H$~~~~~~=~~$z$--coordinate at point of closest approach, 
\vspace*{-4mm}
\item[]\,\,$\cot \theta$~~=~~cotangent of track direction w.r.t. $z$-axis.
%\end{enumerate}
\end{itemize}
Transformations between local parameters (cylindrical or cartesian) 
and global ZEUS perigee parameters are given in appendix D.

%======================= 
 \section{Fast Computations with Kalman Filter Technique} 
%=======================
\label{fastkf}
%Certainly, reduction of an execution time of 
%the track fitting procedure is important to speed up the processing
%of ZEUS data. 
%The majority of computations 
Most of the calculation
by the Kalman filter technique is in the following procedures:
\begin{itemize} 
\item  
transportation and projection of track parameters (\ref{exac}) 
and evaluation of Jacobian matrices (\ref{jakob}); 
\item
matrix operations in prediction (\ref{ck1}),
filter (\ref{kxc}) and smoother (\ref{smooth}) equations;
\item 
search of a track crossing with material to evaluate 
effects of multiple scattering and energy loss. 
\end{itemize}
Approaches to fast computation
with Kalman filter technique were discussed
for the magnet tracking~\cite{magnet},\cite{time05} at the HERA-B detector.

To reduce computations
we use a flexible strategy 
%for track parameters and derivatives propagation 
for propagating track parameters and derivatives
in the inhomogeneous field,
as described for  forward and rear tracks in Subsec.\,\ref{eqmotion}.
For long ($s > 10\,{\rm cm}$) distances
we use numerical integration of the equations of motion, but integrate 
derivatives together with a ``zero trajectory'' 
that allows to reduce computations.
For short distances ($s < 10\,{\rm cm}$) we use 
parabolic expansion (\ref{parabolic}) of the particle trajectory,
which is very fast in computations. 

To keep the computational effort at a minimum
we exploit the sparse structures of the Jacobian matrices.
The Jacobian of track propagation includes elements which are 
very close to 0 or 1, therefore 
we use Jacobian approximations and set such elements
to 0 or 1. 
The Jacobians for cylindrical (\ref{jacob})
and cartesian (\ref{jacobcar}) parameterization
contain only 11 and 10 nontrivial elements, respectively.
To calculate the product of matrices $F_k\,C_{k-1}\,F_k^T$ 
in (\ref{ck1}) 
%page28
we implement functions, which take into
account a sparse structure of the matrix $F_k$.
For example, the function for 10 nontrivial elements of the $F_k$
implies 73 multiplications, which is much smaller than 200
multiplications needed for the case of the completely filled
matrix $F_k$ of size 5 by 5. 

The Jacobians of projection transformation, $H_k$, are
approximated also, 
%by setting trivial elements either to ones or zeroes,
as shown in (\ref{dhdxcylmvd}), (\ref{dhdxcylctd}), (\ref{jacobhcylaxial}),
(\ref{jacobhcylaxial1}) etc.   
We implement corresponding functions for the calculation of products
of $C_k^{k-1}\,H_k^T$ and $(1 - K_k\,H_k)\,C_k^{k-1}$ 
in (\ref{kxc}) or $(1 - H_k\,K_k)\,V_k$ in (\ref{rrk}).
These functions take into account the sparse structure of the matrix $H_k$.
For example, only 20 multiplications are sufficient to 
obtain the matrix $(1 - K_k\,H_k)\,C_k^{k-1}$ for
the option with one nontrivial element in the matrix $H_k$.
This has to be compared with 100 multiplications needed for the
completely filled matrix of size 5 by 1.

To evaluate the 
effects of multiple scattering and energy loss, 
%in (\ref{Qijsum})
we describe the distribution of material in the ZEUS inner trackers 
by using about 
%$1.8\cdot10^3$ 
1800 separate volumes.
After crossing a given volume, a particle can reach only a limited number
of other volumes.
We implement an approach called {\it volume navigation}~\cite{pattern}
for fast search of a track's crossings with these volumes.
Using the Monte Carlo technique,
we evaluate for each volume a list of volumes, which can be
crossed subsequently. 
On average, one list includes about
7 subsequent volumes. 
The lists are used to navigate a fast search of track 
crossing with volumes.  

The described approaches have been programmed~\cite{rtfitexp}
in C++. We follow recipes of effective programming
of numerical calculations~\cite{RK} and implement STL containers
to store objects like hits, states, tracks etc.  

%===========================================================================
%                             Table 1
%===========================================================================
\begin{table}[hbtp]
\begin{center}
\caption{ 
Computing time of the track fit per ZEUS event
on a PC with processor Intel CPU 3.06GHz
for different groups of tracks. 
}
\vspace*{5mm}
\label{cpu}
\begin{tabular}{ c c c c }
\hline 
\hline
 Fitted tracks & Fraction & Field model & Time/event \\
\hline
Forward $(\theta < 60^\circ)$ & $59\%$ & inhomogeneous & 12\,ms \\
Central $(60^\circ< \theta < 120^\circ)$ & $23\%$ & homogeneous & 7\,ms \\
Rear $(\theta > 120^\circ)$ & $18\%$ & inhomogeneous & 1\,ms \\
\hline
All tracks in event& $100\%$  &(in)homogeneous& 20\,ms \\
\hline
\hline
\end{tabular}
\end{center}
\end{table}
%-------------------------------------------------------------------- 
A ZEUS event contains up to 100 fitted tracks and about 30 tracks
on average. The longest tracks include about 80 hits in the central area,
50 hits in a transition region and 30 hits in the very forward direction.
Fitting all the tracks in one event takes 
%on average
$20\,ms$ on PC with processor Intel CPU 3.06GHz (see Table\,\ref{cpu})
and $46\,ms$ with processor Intel CPU 1GHz.
About of $77\%$ of tracks are fitted using the inhomogeneous field
as shown in Table\,\ref{cpu}. The computing time 
%of the fitting 
for these tracks is comparable with those which are fitted 
using the homogeneous field approximation.
%------------------------------------------------------
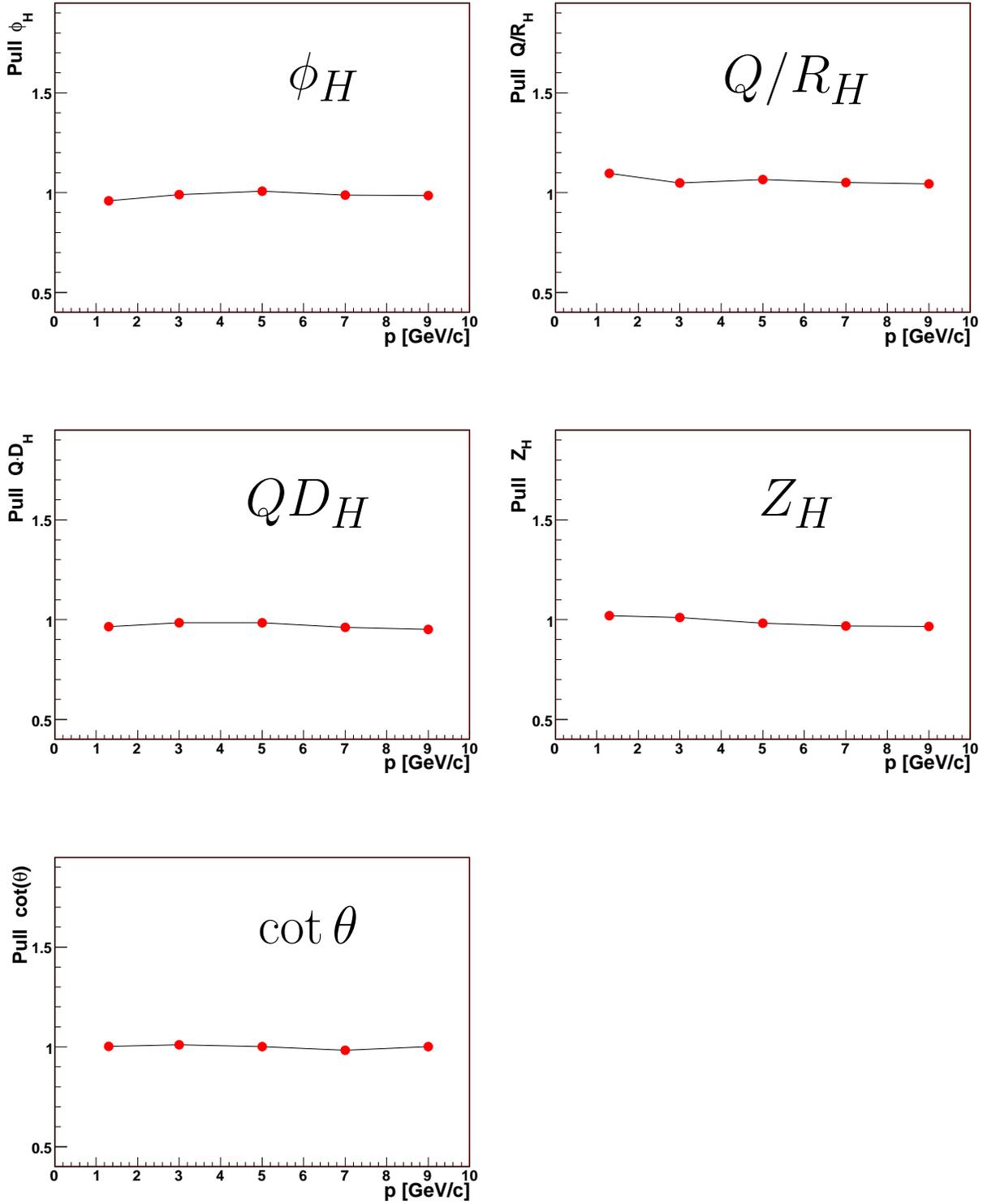
\begin{figure}[hbtp]
\begin{center}
\unitlength1cm
\begin{picture}(15.5,20)
\put( -0.50,14.0){\epsfig
{file=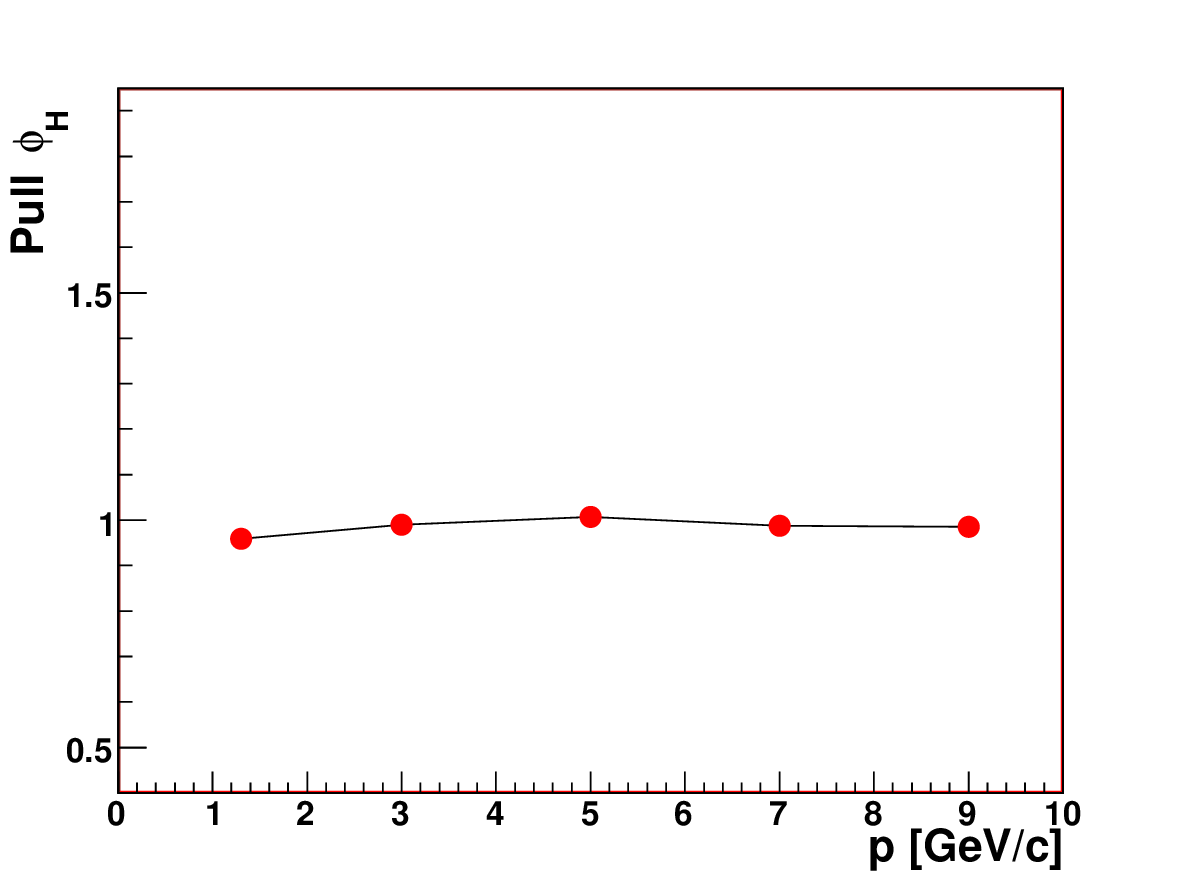,width=8.5cm}}
\put(7.7,14.0){\epsfig
{file=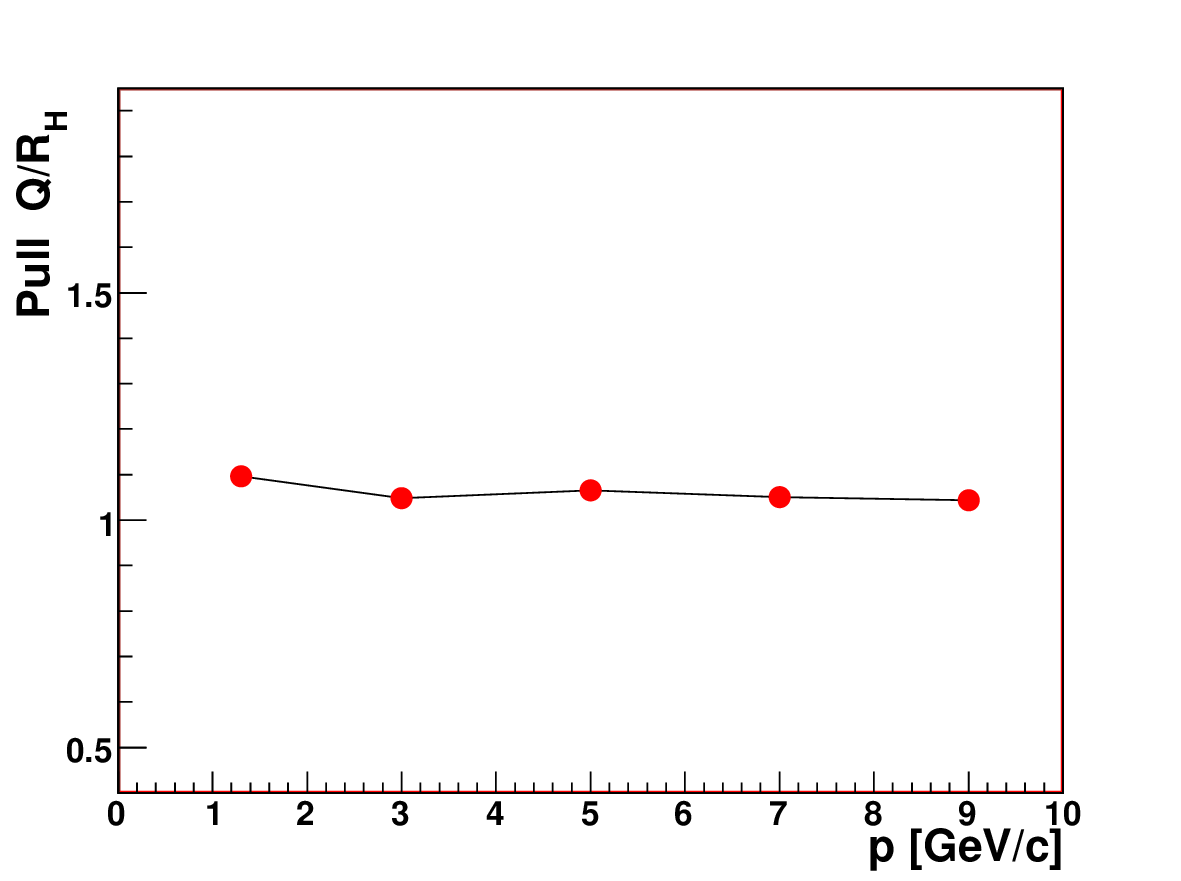,width=8.5cm}}

\put( -0.50,7.0){\epsfig
{file=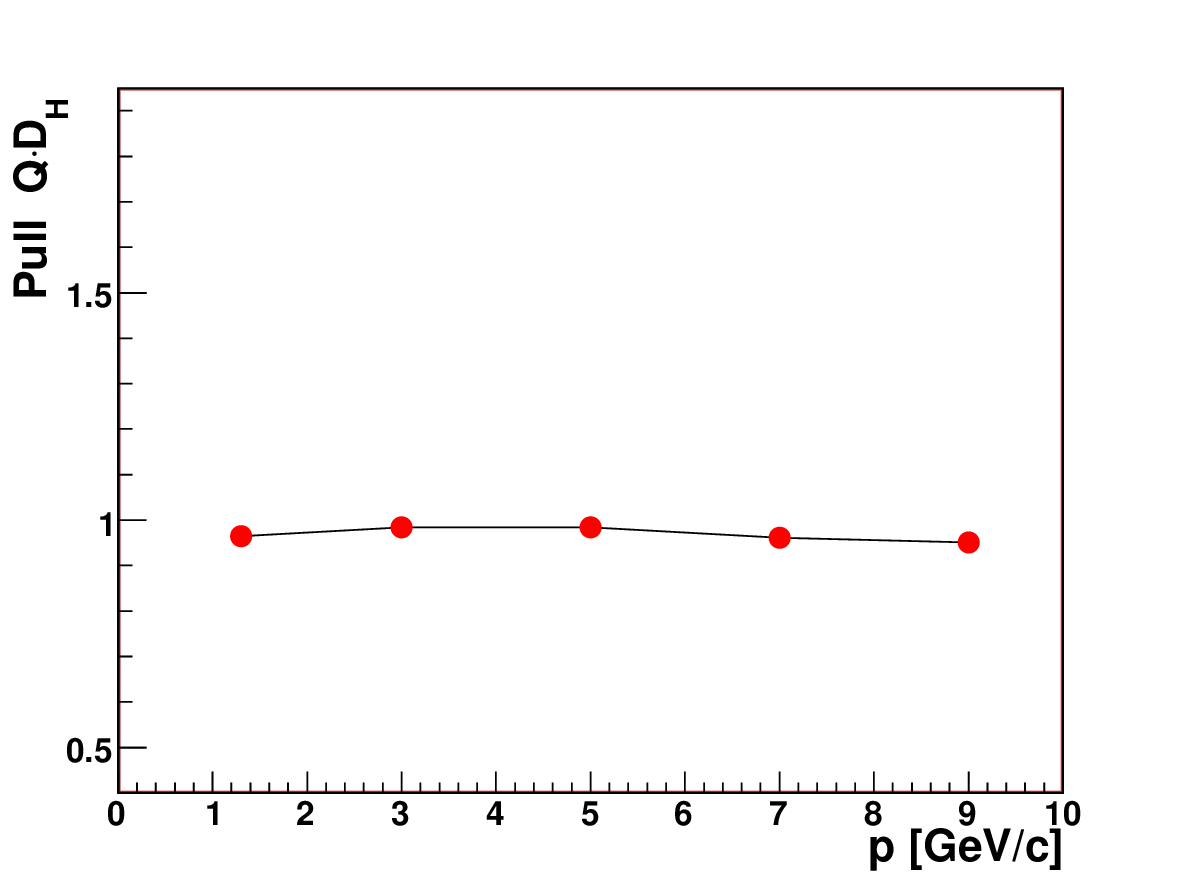,width=8.5cm}}
\put(7.7,7.0){\epsfig
{file=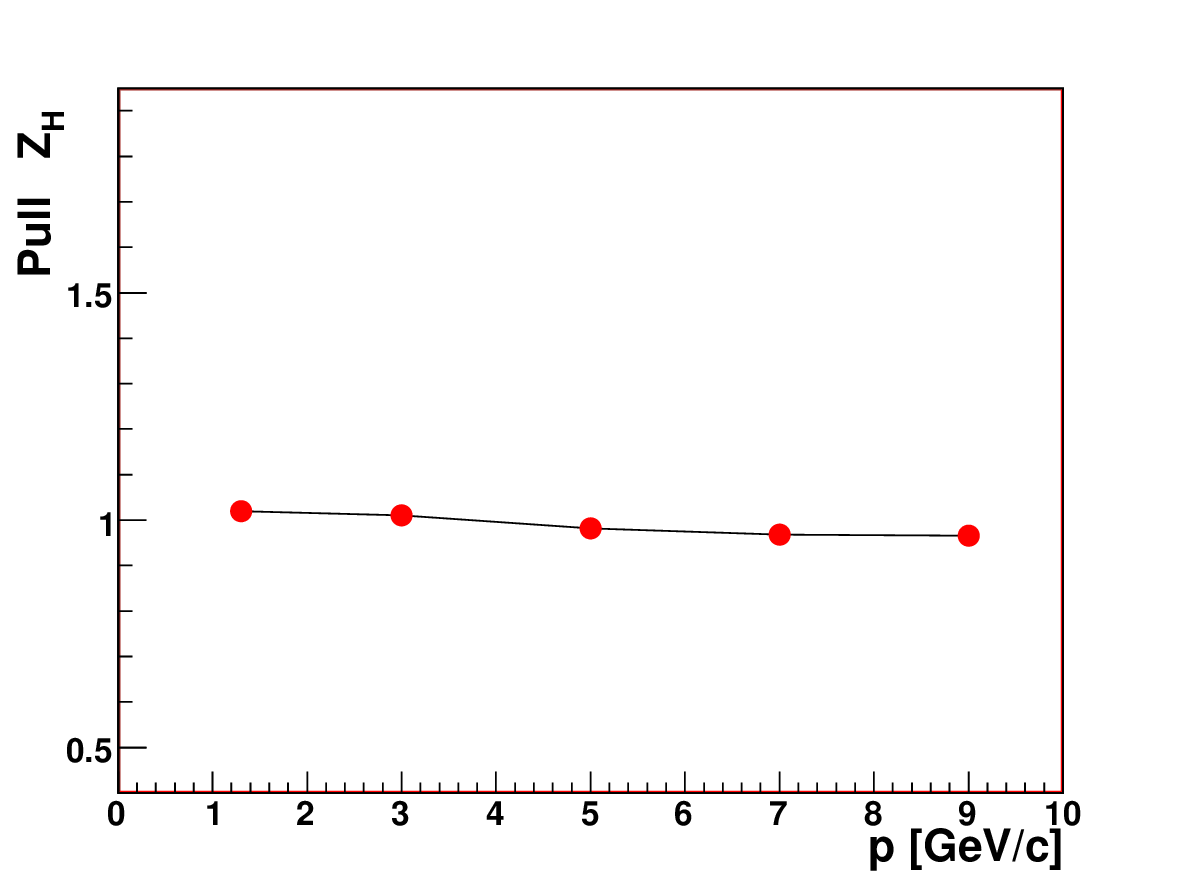,width=8.5cm}}

\put( -0.50,0.0){\epsfig
{file=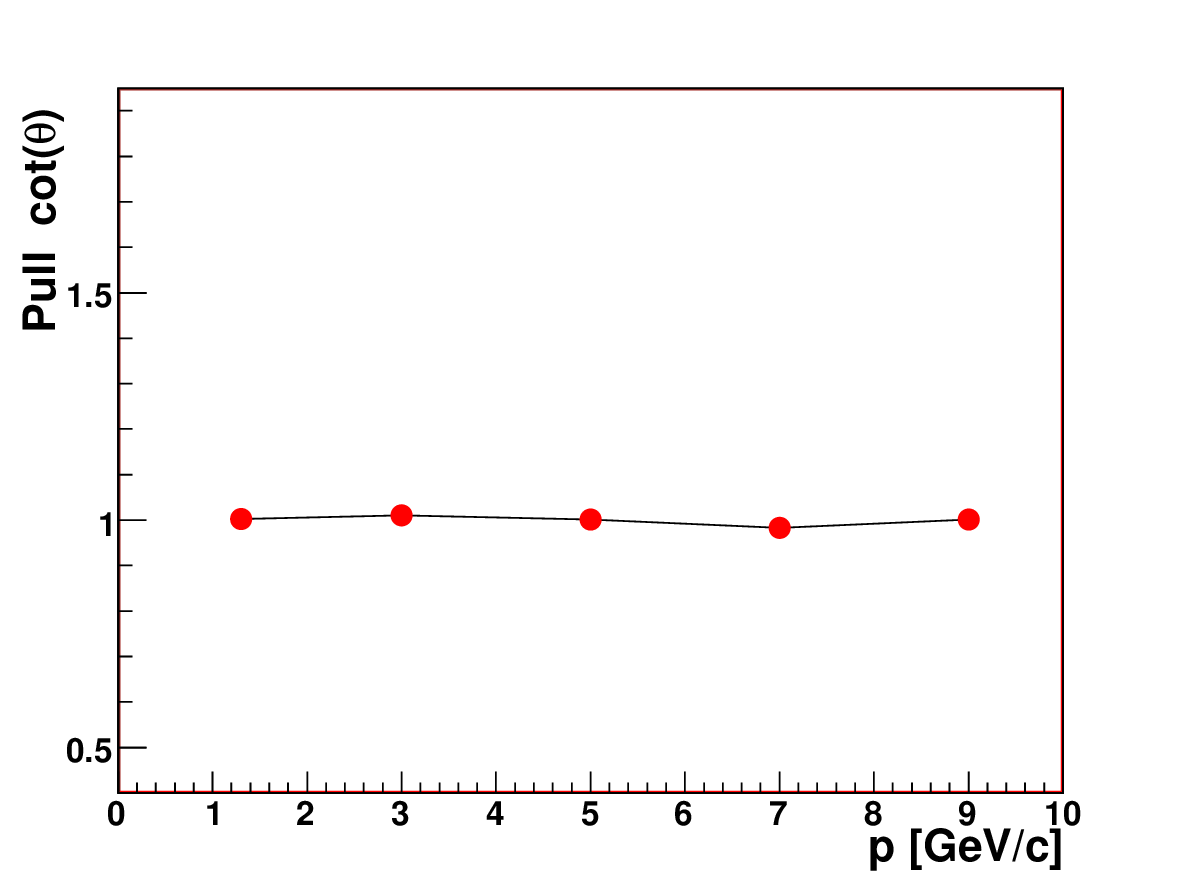,width=8.5cm}}

\put(4.75,18.9){\makebox(0,0)[t]{\huge {$\phi_H$}}}
%\put(2.09,18.9){\makebox(0,0)[t]{\huge SL=5-9}}
\put(12.5,18.9){\makebox(0,0)[t]{\huge $Q/R_H$}}
\put(4.5,11.9){\makebox(0,0)[t]{\huge $QD_H$}}
\put(12.5,11.9){\makebox(0,0)[t]{\huge $Z_H$}}
\put(4.5,4.9){\makebox(0,0)[t]{\huge $\cot \theta$}}
\end{picture}
\caption{\small Standard deviations of the
pull distributions of fitted track parameters in 
perigee parameterization (\ref{xsperigee}) 
for MC simulated muons 
versus the momentum. 
%Results are presented for tracks with CTD outer SL from 5 to 9.
}
\label{pull_mom}
\end{center}
\end{figure}
%--------------------------------------------------------  
\clearpage
%page30
The precision of fitted parameters depends on 
the resolution and details of the performance of the ZEUS trackers and 
will be discussed in the next note~\cite{rtfitexp}.
Here we would like to mention that approximations implemented
to reduce computations, 
are not made at the expense of track parameter precision. 
Evaluation of the covariance matrices of fitted parameters 
in (\ref{kxc}) and (\ref{smooth})
are the most complicated computations,
including operation with the transport, projection and process noise matrices.  
Calculated variances of these matrices are in good agreement
with the residuals of the fitted parameters. Standard deviations
of pull distribution (residuals normalized by their estimated error)
are close to unity for different track momenta, as shown in Fig.\,\ref{pull_mom}. 
%\clearpage
%======================= 
 \section{Conclusions} 
%=======================
We consider a mathematical framework for the rigorous approach to a common 
track fit using the trackers in the inner region of the ZEUS detector:
%Trackers include 
CTD, BMVD, FMVD and STT.
We discuss track models and likelihood functions in such a multi-component
tracker.
The approach offers a rigorous treatment of field inhomogeneity,
multiple scattering and energy loss.
The track fitting procedure makes use 
of the Kalman filter technique.

We describe details of the mathematics
for the fast implementation of a Kalman filter for
the cylindrical drift chamber, barrel and forward silicon strip detectors
and straw drift chambers. The cases of homogeneous and inhomogeneous
field are considered.  

We discuss how to reduce computations
and make the track fitting procedure fast.
Average computing time of track fitting in one ZEUS event
is about of 20\,ms on a PC with processor Intel CPU 3.06GHz.

\vspace{6mm}
%\newline
{\bf Acknowledgments:}
%=====================
%\section{Acknowledgments}
%=====================
Fruitful discussions with R.\,Mankel helped me 
a lot to finalize the results. 
%I am also grateful to him 
%for the careful reading of the manuscript. 
Cordial thanks to A.\,Antonov, C.\,Catteral and G.\,Hartner 
%and V.\,Roberfroid 
for their expertise on the implementation of ZEUS trackers
in the simulation and reconstruction
software.
I am grateful to O.\,Behnke and G.\,Hartner 
for the careful reading of the manuscript. 
I would like to thank the ZEUS group at DESY for the kind hospitality 
extended to me during my visit. 

%I had great benefit from the 
%expertise and advices of F.\,Jegerlehner
%whose detailed analytic results for radiative decays of Z bosons were
%the starting point for this study. 
%Cordial thanks to T.\,Lohse who stressed the importance of
%radiative decays for charmonia measurements and developed 
%the Monte Carlo generator used for the investigation.
%Fruitful discussions with G.\,Bohm and H.\,Kolanoski helped me
%a lot to finalize the results. I am also grateful to them
%for the careful reading of the manuscript.
%I would like to thank DESY Zeuthen for the kind hospitality extended 
%to me during my visit.
%==================================================== References
%
%\newpage
%

%page32
\clearpage
%======================= 
 \section{Appendix A: Jacobian of prediction transformation in cylindrical 
parameterization}
%=======================
We use derivatives of $t_k$ 
to calculate the Jacobian (\ref{jacob}) 
of prediction transformation (\ref{xk1}):
%with respect to componets of the ${x}_k$: 
\begin{equation}
\begin{array}{ll}
\partial t_k/\partial u_k & = 
\frac {\displaystyle \cos(\phi_k - \frac{u_k}{r_k})
- \cos(\phi_k - \frac{u_k}{r_k} - t_k)}
{\displaystyle r_k\, \cos(\phi_k - \frac{u_k}{r_k} - t_k)
+ \frac{1}{\omega_k} \,\sin t_k }, \\
\partial t_k/\partial z_k & = 0, \\ 
\partial t_k/\partial \phi_k & = -r_k\,\,\partial t_k/\partial u_k, \\

\partial t_k/\partial \lambda_k & = \displaystyle \frac{\omega_k \lambda_k}
{1 + \lambda_k^2}\,\partial t_k/\partial \omega_k, \\ 

\partial t_k/\partial \omega_k & = 
\frac {\displaystyle 2\,(1-\cos t_k) + r_k\, \omega_k 
\left[\sin (\phi-\frac{u_k}{r_k}) 
- \sin (\phi-\frac{u_k}{r_k} - t_k) \right]  }
{\displaystyle r_k\,\omega_k^2\,\cos (\phi-\frac{u_k}{r_k} - t_k)
+ \omega_k\,\sin t_k }.~~~~~~~~~
\end{array} 
\label{dtdx}     
\end{equation} 
Nontrivial elements of the Jacobian are: 
\begin{equation} 
\begin{array}{ll}
\partial u_{k+1}/\partial u_k & 
=\,\, \frac{\displaystyle r_{k+1}}{\displaystyle x_{k+1}} \left[
\cos \frac{u_k}{r_k} + \frac{1}{\omega_k} \sin (\phi_k - t_k)\,
\partial t_{k}/\partial u_k  \right]~~~~~~~~~~~~~~~~~~~~~~~\\
 & \\
 & =\,\, \frac{\displaystyle r_{k+1}}{\displaystyle y_{k+1}} \left[
\sin \frac{u_k}{r_k} - \frac{1}{\omega_k} \cos (\phi_k - t_k)\,
\partial t_{k}/\partial u_k  \right],\\
 & \\
\partial u_{k+1}/\partial \phi_k & 
= \frac{\displaystyle r_{k+1}}{\displaystyle \omega_k\,x_{k+1}} \left[
\sin \phi_k - \sin (\phi_k - t_k)\,
(1 - \partial t_{k}/\partial \phi_k)  \right]\\
 & \\
 & 
=\,\, \frac{\displaystyle r_{k+1}}{\displaystyle \omega_k\,y_{k+1}} \left[
-\cos \phi_k + \cos (\phi_k - t_k)\,
(1 - \partial t_{k}/\partial \phi_k)  \right],\\
 & \\
\partial u_{k+1}/\partial \lambda_k & 
= \frac{\displaystyle r_{k+1}}{\displaystyle \omega_k\,x_{k+1}} \left[
-\dvrat{\lambda_k}{1 + \lambda_k^2}
\left( \cos (\phi_k - t_k)-\cos \phi_k \right) 
+\sin (\phi_k - t_k)\,\partial t_{k}/\partial \lambda_k  \right]\\
 & \\
 & 
= \frac{\displaystyle r_{k+1}}{\displaystyle \omega_k\,y_{k+1}} \left[
-\dvrat{\lambda_k}{1 + \lambda_k^2}
\left( \sin (\phi_k - t_k)-\sin \phi_k \right) 
-\cos (\phi_k - t_k)\,\partial t_{k}/\partial \lambda_k  \right],\\
 & \\
\partial u_{k+1}/\partial q_k & 
=\,\, \frac{\displaystyle r_{k+1}}{\displaystyle q_k\,\omega_k\,x_{k+1}} \left[
\cos \phi_k - \cos (\phi_k - t_k)
+\omega_k\,\sin (\phi_k - t_k)\,\partial t_{k}/\partial \omega_k \right]\\
 & \\
& =\,\,\frac{\displaystyle r_{k+1}}{\displaystyle q_k\,\omega_k\,y_{k+1}}\left[
\sin \phi_k - \sin (\phi_k - t_k)
-\omega_k\,\cos (\phi_k - t_k)\,\partial t_{k}/\partial \omega_k \right],\\
 & \\
\partial z_{k+1}/\partial u_k &
=\,\, \frac{\displaystyle \lambda_k} {\displaystyle \omega_k}\,
\partial t_{k}/\partial u_k,~~~~~~~~~~~
\partial z_{k+1}/\partial \phi_k 
=\,\, \frac{\displaystyle \lambda_k} {\displaystyle \omega_k}\,
\partial t_k/\partial \phi_k,\\
% & \\
\partial z_{k+1}/\partial \lambda_k &
=\,\, \frac{\displaystyle t_k} {\displaystyle \omega_k},~~~~~~~~~~~~~~~~~~~~~~
\partial z_{k+1}/\partial q_k 
=\,\, \frac{\displaystyle \lambda_k} {\displaystyle q_k}\,
\left( \partial t_k/\partial \omega_k - 
\frac{\displaystyle t_k} {\displaystyle \omega_k} \right),\\
% & \\
\partial \phi_{k+1}/\partial u_k &
=\, -\partial t_k/\partial u_k,~~~~~~~~~~~~~
\partial \phi_{k+1}/\partial \lambda_k\, 
=\,\, \displaystyle\frac{\lambda_k \omega_k}{1 + \lambda_k^2}
\partial t_k/\partial \omega_k,\\
\partial \phi_{k+1}/\partial q_k &
=\,\,-\partial t_k/\partial q_k. \\
\end{array}  
\label{dxk1dxk}      
\end{equation}
   
%page33
%======================= 
 \section{Appendix B: Jacobian of projection transformation for 
the CTD in cylindrical parameterization}
%=======================
Elements of the Jacobian (\ref{dhdxcylctd}) 
of projection transformation (\ref{hkcylctd})
for the stereo CTD are:
\begin{equation}
\begin{array}{ll} 
\partial h_k / \partial z_k& = {\cal C}_k\,\partial \Delta t_k/\partial z_k
- m_x\,r'_{wx} - m_y\,r'_{wy},\\
\partial h_k / \partial \phi_k& = 
{\cal C}_k\,\partial \Delta t_k/\partial \phi_k
+ \frac{\displaystyle \Delta t_k}{\displaystyle \omega_k}
\left( -m_x\,\sin\phi_k + m_y\,\cos\phi_k \right), \\
\partial h_k / \partial \lambda_k& = 
{\cal C}_k\,\partial \Delta t_k/\partial \lambda_k
-\Delta t_k \left[ \frac{\displaystyle \lambda_k\,{\cal C}_k}
{\displaystyle 1 + \lambda_k^2}
+ \frac{\displaystyle 1}{\displaystyle \omega_k}
\left( m_x\,r'_{wx} + m_y\,r'_{wy} \right)  \right],\\
\end{array}  
\label{dhdxcylctd1}      
\end{equation}   
with derivatives
\begin{equation}
\begin{array}{ll} 
\partial\Delta t_k/\partial z_k&=  
-\frac{\displaystyle \Delta t_k\,\partial b_k/\partial z_k 
+ \partial c_k/\partial z_k}
{\displaystyle 2\,\Delta t_k\,a_k + b_k},\\
\partial\Delta t_k/\partial \phi_k&= 
-\frac{\displaystyle \Delta t_k\,\partial b_k/\partial \phi_k}
{{\Delta t_k}^2\,\displaystyle \partial a_k/\partial \phi_k + 
2\,\Delta t_k\,a_k + b_k}, \\ 
\partial\Delta t_k/\partial \lambda_k&=  
-\frac{\displaystyle \Delta t_k\,\partial b_k/\partial \lambda_k
+ \partial c_k/\partial \lambda_k}
{{\Delta t_k}^2\,\displaystyle \partial a_k/\partial \lambda_k +  
2\,\Delta t_k\,a_k + b_k},
\\
~~\partial b_k/\partial z_k&=
\,\,\frac{\displaystyle \omega_k}{\displaystyle \lambda_k}\,
\left(A_{kx}\,p'_{wx} + A_{ky}\,p'_{wy} \right)
- p'_{wx}\,r'_{wx} - p'_{wy}\,r'_{wy},~~~~~~~~~~~ \\
~~\partial c_k/\partial z_k&=  
\,\,\frac{\displaystyle \omega_k}{\displaystyle \lambda_k}\,
\left(B_{kx}\,p'_{wx} + B_{ky}\,p'_{wy} \right)
- {\cal P}_{kx}\,r'_{wx} - {\cal P}_{ky}\,r'_{wy}, \\
~~\partial a_k/\partial \phi_k&=
\,\,\frac{\displaystyle 1}{\displaystyle \omega_k}\,
\left(-\sin\phi_k\,p'_{wx} + \cos\phi_k\,p'_{wy} \right), \\
~~\partial b_k/\partial \phi_k&=
\,\,\frac{\displaystyle 1}{\displaystyle \omega_k}\,
\left(-\sin\phi_k\,{\cal P}_{kx} + \cos\phi_k\,{\cal P}_{ky} \right), \\
~~\partial a_k/\partial \lambda_k&=
\,\,p'_{wx}\,\partial A_{kx}/\partial \lambda_k
+ p'_{wy}\,\partial A_{ky}/\partial \lambda_k, \\
~~\partial b_k/\partial \lambda_k&= 
\,\,{\cal P}_{kx}\,\partial A_{kx}/\partial \lambda_k 
+ {\cal P}_{ky}\,\partial A_{ky}/\partial \lambda_k\\
&
~~~-\frac{\displaystyle 1}{\displaystyle \lambda_k\,(1 + \lambda_k^2)}
\left( A_{kx}\,{\cal P}_{kx} + A_{ky}\,{\cal P}_{ky} \right),\\
~~\partial c_k/\partial \lambda_k&=
-\frac{\displaystyle 1}{\displaystyle \lambda_k\,(1 + \lambda_k^2)}
\left( B_{kx}\,{\cal P}_{kx} + B_{ky}\,{\cal P}_{ky} \right)
,\\
~~\partial A_{kx}/\partial \lambda_k&=
-\frac{\displaystyle 1}{\displaystyle \omega_k}
\left[ \frac{\displaystyle \lambda_k}
{\displaystyle 1 + \lambda_k^2}
(\cos \phi_k - \lambda_k r'_{wx}) + r'_{wx} \right],\\
~~\partial A_{ky}/\partial \lambda_k&=
-\frac{\displaystyle 1}{\displaystyle \omega_k}
\left[ \frac{\displaystyle \lambda_k}
{\displaystyle 1 + \lambda_k^2}
(\sin \phi_k - \lambda_k r'_{wy}) + r'_{wy} \right].\\
\end{array}  
\label{dtdxcylctd}      
\end{equation}   
Elements of the corresponding Jacobian (\ref{jacobhcylaxial}) for the axial CTD 
look as:
\begin{equation}
\begin{array}{ll} 
\hder{h_{k1}}{\phi_k}& = 
\dvrat{m_{wx}}{\omega_k}\,(-\Delta t_k \sfk + \dvder{\Delta t_k}{\phi_k}\,\cfk)
\\
&~~~~~~~~~~~~~~~~~~~~~~~~~
+ 
\dvrat{m_{wy}}{\omega_k}\,(\Delta t_k \cfk + \dvder{\Delta t_k}{\phi_k}\,\sfk)
,\\
\hder{h_{k2}}{u_k}& =
\dvrat{\lambda_k}{a_k\,\omega_k}
(p_{wx}\dvrat{y_k}{r_k} - p_{wy}\dvrat{x_k}{r_k}), \\
\\
\hder{h_{k2}}{\phi_k}& =
\dvrat{\lambda_k}{\omega_k}\,\dvder{\Delta t_k}{\phi_k},\\
\\
\hder{h_{k2}}{\lambda_k}& = \dvrat{\Delta t_k}{\omega_k},\\
\\
\hder{\Delta t_k}{\phi_k}& =
\dvrat{\Delta t_k}{a_k\,\omega_k}
(p_{wx}\,\sfk - p_{wy}\,\cfk). \\

\end{array}
\label{dhdxcylaxial}      
\end{equation}

%page34
%======================= 
 \section{Appendix C: Jacobian of projection transformation for 
the CTD in cartesian parameterization}
%=======================
Nontrivial elements of the Jacobian (\ref{hcarctds}) 
of projection transformation (\ref{hkcylctdcar})
for the stereo CTD are: 
\begin{equation} 
\begin{array}{ll} 
\hder{h_k}{x_k} = m_x + M_w\,\dvder{\Delta s_k}{x_k}, & 
~~~~~~~~\hder{h_k}{y_k} = m_y + M_w\,\dvder{\Delta s_k}{y_k}, \\ 
\hder{h_k}{t_{xk}} = \dvder{h_k}{x_k}\cdot\Delta s_k, & 
~~~~~~~~\hder{h_k}{t_{yk}} = \dvder{h_k}{y_k}\cdot\Delta s_k, \\ 
\end{array}   
\label{dhdxcarctds}       
\end{equation}    
with 
\[
 \hder{\Delta s_k}{x_k} = 
\displaystyle -\dvrat{\Delta s_k\,p'_{wx}+{\cal P}_{kx}}
	{2\,\Delta s_k\,a_k+b_k},~~~~~~
 \hder{\Delta s_k}{y_k} = 
\displaystyle -\dvrat{\Delta s_k\,p'_{wy}+{\cal P}_{ky}}
	{2\,\Delta s_k\,a_k+b_k}, 
\]
\[
M_w = m_{wx}\,(t_{kx}-r'_{wx}) + m_{wy}\,(t_{ky}-r'_{wy}).~~~~~~~~~~~~~~~~~~~~~~~~~~~~~~~~
\]
Elements of the corresponding Jacobian (\ref{jacobcaraxial}) for the axial CTD
read as: 
\begin{equation} 
\begin{array}{ll} 
\displaystyle
\hder{h_{k1}}{x_k} = m_x + M_w\,\dvder{\Delta s_k}{x_k}, & 
~~~~~~~~\dvder{h_{k1}}{y_k} = m_y + M_w\,\dvder{\Delta s_k}{y_k}, \\ 
\displaystyle 
\hder{h_{k1}}{t_{xk}} = \dvder{h_{k1}}{x_k}\cdot\Delta s_k, & 
~~~~~~~~\dvder{h_{k1}}{t_{yk}} = \dvder{h_{k1}}{y_k}\cdot\Delta s_k, \\ 
\displaystyle 
\hder{h_{k2}}{x_k} = \dvder{\Delta s_k}{x_k}, & 
~~~~~~~~\dvder{h_{k2}}{y_k} = \dvder{\Delta s_k}{y_k}, \\ 
\displaystyle 
\hder{h_{k2}}{t_{xk}} = \dvder{\Delta s_k}{x_k}\cdot\Delta s_k, & 
~~~~~~~~\dvder{h_{k2}}{t_{yk}} = \dvder{\Delta s_k}{y_k}\cdot\Delta s_k, \\ 

\end{array}   
\label{dhdxcarctda}       
\end{equation}    
with 
\[
 \hder{\Delta s_k}{x_k} = 
\displaystyle -\dvrat{p_{wx}}{a_k},~~~~~~~~~~~~~~~~~~~
 \hder{\Delta s_k}{y_k} = 
\displaystyle -\dvrat{p_{wy}}{a_k},~~~~~~~ 
\]
\[
M_w = m_{wx}\,t_{kx} + m_{wy}\,t_{ky}.~~~~~~~~~~~~~~~~~~~~~~~~~~~~~~~~~~~~~~~~~~~~~~~~~~
\]

%======================= 
 \section{Appendix D: Conversions from Local to Global Parameters}
%=======================
Track parameters, $u_0, z_0, \phi_0, \lambda_0, q_0$, at the beginning 
of a central track, which is fitted using 
the cylindrical parameterization (\ref{xsol}), are converted to perigee 
parameters (\ref{xsperigee}):
\begin{equation}
\begin{array}{ll}
 \phi_H&=~~\phi_0 - t_H,\\
 Q/R_H&=~~\kappa\,B_{z}\,q_0\,\sqrt{1 + \lambda_0^2}, \\ 
 QD_H&=~~\displaystyle -r_0 \sin(\dvrat{u_0}{r_0} - \phi_0 + t_H) 
+ (\cos t_H - 1)\,/\,(Q/R_H),\\	
 z_H& =~~z_0 + \lambda_0\,t_H\,/\,(Q/R_H),\\
 \cot \theta&=~~\lambda_0,\\	
\end{array} 
\label{cylper}     
\end{equation} 
where
\[
\begin{array}{ll}
~~~~~~~t_H&=~~\arctan  \left[
\dvrat{ 1/(Q/R_H) - r_0\,\sin (u_0/r_0 
- \phi_0)}{r_0\,\cos (u_0/r_0 - \phi_0)} \right]
 - \dvrat{\pi}{2}\,{\rm sign}(Q/R_H).
\end{array}
\]
%page35
We convert 
fitted cartesian parameters 
%(\ref{xf}) 
at the beginning of a forward track
,$x_0, y_0, t_{0x}, t_{0y}, q_0$,
into perigee parameters (\ref{xsperigee}):
\begin{equation}
\begin{array}{ll}
 \phi_H&=~~\displaystyle \arctan \dvrat{{\cal Y}_0}{{\cal X}_0},\\
 Q/R_H&=~~\displaystyle \kappa\,B_{z}\,q_0\,
\dvrat{\sqrt{1 + t_{0x}^2 + t_{0y}^2}}{\sqrt{t_{0x}^2 + t_{0y}^2}}, \\ 
 QD_H&=~~\displaystyle \frac{-1 + {\cal Y}_0\,\sin \phi_H 
	+ {\cal X}_0\,\cos \phi_H}{Q/R_H},~~~~~~~~~~~~\\
 z_H& =~~\displaystyle z_0 + \dvrat{\Phi_0 - \phi_H}
{\sqrt{t_{0x}^2 + t_{0y}^2}\,\,Q/R_H},\\
 \cot \theta_H&=~~\displaystyle \dvrat{1}{\sqrt{t_{0x}^2 + t_{0y}^2}},\\
\end{array} 
\label{carper}     
\end{equation} 
where
\[
\begin{array}{ll}
\Phi_0&=~~\displaystyle \arctan \dvrat{t_{0y}}{t_{0x}},\\
{\cal X}_0&=\displaystyle -y_0\,Q/R_H 
	+ \frac {t_{0x}}{\sqrt{t_{0x}^2 + t_{0y}^2}},~~~~~~~~~~~~~~~~~\\
{\cal Y}_0&=~~\displaystyle x_0\,Q/R_H 
	+ \frac {t_{0y}}{\sqrt{t_{0x}^2 + t_{0y}^2}}.\\
\end{array}
\]
The transformation from cartesian to perigee parameterization
for rear tracks is similar to (\ref{carper}) for parameters
$\phi_H, Q/R_H, QD_H$, but differs for parameters $z_H$ and $\cot \theta_H$:
\begin{equation}
\begin{array}{ll}
 z_H& =~~\displaystyle z_0 - \dvrat{\Phi_0 - \phi_H}
{\sqrt{t_{0x}^2 + t_{0y}^2}\,\,Q/R_H},~~~~~~~~~~~~\\
 \cot \theta_H&=~~-\displaystyle \dvrat{1}{\sqrt{t_{0x}^2 + t_{0y}^2}}.\\
\end{array} 
\label{cabper}     
\end{equation} 

\end{document}